# Dynamic Modeling and Control of Phosphate-Pebble Drying Systems−A Comprehensive Approach

José M. Campos-Salazar
*Electronic Engineering Department*
*Universitat Politècnica de Catalunya*
Barcelona, Spain
jose.manuel.campos@upc.edu

Felipe Santander
*Process Engineering Department*
*Celulosa Arauco y Constitución, SA*
Concepción, Chile
felipe.santander@arauco.com

Eduardo Keim
*Process Research Department*
*Bioforest, SA*
Concepción, Chile
eduardo.keim@arauco.com

***Abstract*—Dryers play a central role in the processing of phosphate rock, where moisture removal is essential for downstream handling and energy efficiency. Due to the inherently nonlinear and multivariable nature of these systems, accurate modeling and control remain industrial challenges. This article presents a comprehensive nonlinear dynamic model of a phosphate-pebble rotary drying process, built from first principles to capture coupled heat and mass transfer, evaporation kinetics, and subsystem interactions. A multivariable control strategy is developed using direct synthesis–based tuning of decentralized PI compensators to regulate moisture content, combustion temperature, and exhaust draft. The model is used to evaluate closed-loop performance and simulate transient responses under realistic disturbances. Additionally, sensitivity analysis and thermal efficiency modeling provide insight into energy utilization. The results offer a unified framework for dynamic analysis, control design, and operational optimization in phosphate drying systems.**

***Keywords—Nonlinear dynamics; phosphate drying; pi control; thermal efficiency***

## I. Introduction

Drying operations are fundamental, yet energy-intensive unit processes extensively deployed across chemical, mineral, and agri-food industries. In the context of phosphate fertilizer production, mined phosphate "pebbles" must undergo thermal dehydration to reduce their inherent moisture content prior to acidulation and downstream processing. This drying step is driven by two pivotal industrial objectives: (i) minimizing the transport of excess water, which translates to reduced logistical costs, and (ii) improving the flowability, grindability, and overall handling characteristics of the solid feedstock [1]. Industrial phosphate dryers are typically large-scale rotary drums wherein the wet solids are exposed to high-temperature combustion gases, facilitating concurrent convective heat and mass transfer. These systems exhibit complex nonlinear dynamics arising from the interaction between thermal and moisture transport phenomena, spatial and temporal transport delays, and the coupled behavior of the gas and solid phases. Consequently, accurate dynamic modeling is essential not only for process design and optimization but also for control synthesis aimed at maintaining product quality and energy efficiency under transient conditions.

Several modeling paradigms have been developed to represent drying systems, ranging from empirical correlations and black-box system identification to high-fidelity first-principles and computational fluid dynamics (CFD) approaches. Empirical thin-layer models, such as the Newton, Page, and Henderson–Pabis formulations [2], remain widely used in laboratory-scale studies owing to their simplicity and ease of parameterization. While these models capture moisture-removal trends over time using one or more adjustable parameters, their reliance on curve-fitting techniques and limited physical insight renders them unsuitable for extrapolation beyond calibrated operating conditions. Despite these limitations, they continue to be cited in the literature for a wide variety of agricultural and food products due to their minimal data requirements and acceptable predictive accuracy under fixed settings [2].

In contrast, black-box system identification and machine-learning-based techniques have gained traction for modeling complex drying systems where first-principles derivation is challenging. Approaches including ARX/ARMAX models, support vector regression, and adaptive neuro-fuzzy inference systems have been applied to rotary dryers with promising accuracy [3], [4]. However, the predictive capability of these models is typically confined to the training domain, and their interpretability is inherently limited. Consequently, such models are often integrated with physics-based approaches when control-oriented modeling is required.

CFD offers a high-resolution alternative for simulating coupled heat, momentum, and species transport within dryers. State-of-the-art studies employ multiphysics CFD solvers coupled with discrete element methods (DEM) to resolve solid–gas interactions in packed beds or rotary drums [5]. Examples include the 3D modeling of iron ore pellet dryers and the simulation of laboratory-scale rotary kilns incorporating radiation and detailed kinetics [6], [7]. These studies provide deep insight into localized phenomena such as velocity fields, temperature gradients, and internal diffusion. However, the computational expense, complexity of mesh generation, and parameter calibration significantly limit the practicality of CFD for dynamic control studies or real-time applications.

To bridge the gap between empirical simplicity and CFD detail, first-principles dynamic models based on conservation laws (mass, energy) have emerged as an effective compromise. These models offer physically grounded formulations capable of capturing drying kinetics, gas–solid heat exchange, and internal moisture transport, often with reduced computational burden.

The literature documents several seminal contributions in this area, such as the lumped and distributed models for rotary dryers developed in [8]–[10], which laid the foundation for simulating transient drying behavior under various operating scenarios. More recent examples include dynamic models validated against semi-industrial dryers for citrus residues and soybean seeds [11], [12], confirming the utility of this approach in both simulation and control design. Notably, such models yield systems of ordinary or partial differential equations that can be linearized for model-based controller synthesis.

Despite the maturity of drying-system modeling in general, a critical knowledge gap exists regarding phosphate pebble drying systems. Specifically, no existing study reports a comprehensive dynamic model encompassing the full process—from combustion and hot gas generation to the drying drum and exhaust subsystems—for phosphate rock dryers. This is surprising given the unique characteristics of phosphate drying: dense, heterogeneous solids with high thermal inertia and internal moisture gradients, large flow rates, and dusty exhaust environments. Prior works have been either steady-state in nature or partial in scope, focusing on isolated subsystems such as the furnace or airflow control [4], [13]. This lack of a unified model hinders the development of effective control strategies for energy efficiency and product quality in phosphate drying operations.

The present study addresses this deficiency by introducing a novel nonlinear dynamic model of an industrial phosphate-pebble drying system. The model is formulated from first principles and integrates all critical components: (i) the combustion subsystem (fuel and air mixing, burner dynamics), (ii) the rotary drying drum (with gas–solid energy and mass balances), and (iii) the exhaust gas handling unit (including stack flow and draft pressure). The drying kinetics account for both external convective transfer and internal moisture diffusion, enabling accurate simulation of the transition from constant-rate to falling-rate drying regimes. The resulting model comprises a coupled set of nonlinear differential equations, capturing the dynamic propagation of disturbances through the system. To the best of our knowledge, this is the first time such a comprehensive dynamic framework has been proposed and validated for phosphate drying.

Building on this model, the study further develops a multivariable control strategy using decentralized PI compensators tuned via the direct synthesis method (DSM). The linearized model around a nominal operating point facilitates the derivation of decoupled controllers with targeted performance metrics.

Simulation results under multiple disturbance scenarios demonstrate superior setpoint tracking, reduced overshoot, and robust disturbance rejection—highlighting the efficacy of the proposed control approach relative to conventional tuning heuristics.

In summary, this article makes two key contributions. First, it introduces the first comprehensive, first-principles dynamic model of an industrial phosphate pebble dryer that integrates combustion, drying, and exhaust dynamics. Second, it proposes and evaluates a DSM-based multivariable control strategy suitable for real-world deployment. The remainder of the article presents the model formulation, control system design, simulation results, sensitivity analysis, and concluding insights. Together, these contributions aim to fill a significant gap in the literature and provide actionable guidance for the optimization and control of phosphate drying processes.

The remainder of the paper is organized as follows. Section 2 outlines the industrial phosphate dryer system. Section 3 details the dynamic model formulation. Section 4 presents model validation and sensitivity analysis. Section 5 describes the control strategy and tuning. Section 6 covers implementation aspects. Section 7 discusses simulation results, and Section 8 concludes the study.

## II. Process Description

The simplified process flow diagram of the phosphate-pebble dryer is shown in Fig. 1. As illustrated, the dryer is composed of four primary subsystems: (i) the combustion furnace, (ii) the windbox, (iii) the drying zone (pebble bed), and (iv) the exhaust stack. Each subsystem acts as a thermodynamic control volume, where mass, energy, and momentum balances govern the transient process dynamics. Together, they enable the conversion of fuel chemical energy into sensible heat, its transfer into the solid phase, and the removal of evaporated moisture from the phosphate matrix.

### *A. Conbustion Furnace*

The combustion furnace is the primary source of thermal energy. It receives two inlet streams:

- A fuel stream $\dot{m}_{\mathrm{f}}(t)$, typically oil or natural gas, delivered by a pump.
- A combustion air stream $\dot{m}_{\mathrm{a}}(t)$, supplied by a forced-draft fan (FD-fan).

Within the furnace, complete combustion takes place, converting the chemical energy of the fuel into hot flue gases. The composition of these gases is dominated by nitrogen, carbon dioxide, and water vapor. The furnace maintains high gas temperatures, typically between 900–1200 °C, to provide a strong driving force for downstream heat transfer. The outlet stream of the furnace is the hot flue gas flow directed into the windbox. The design of this stage ensures stable combustion,

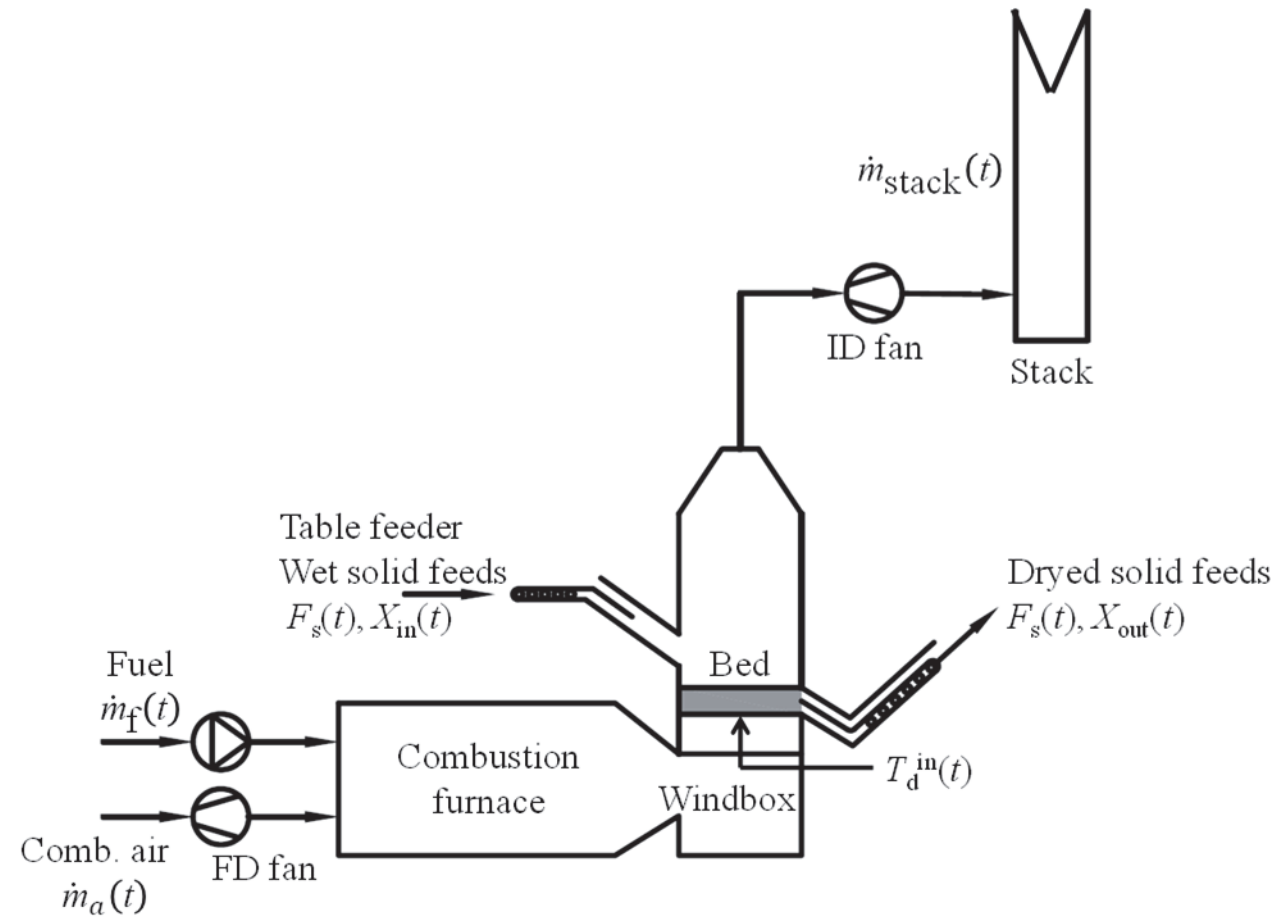

Fig. 1. Phosphate-pebble drying systems. Process diagram.

minimal excess oxygen, and proper mixing of air and fuel for consistent operation [14]–[16].

### B. Windbox

The windbox acts as a plenum chamber between the furnace and the drying zone. Its purpose is to equalize pressure fluctuations, stabilize flow, and uniformly distribute combustion gases across the bed of pebbles.

The incoming furnace gas enters the windbox at high temperature and is conditioned before reaching the bed. The key output of this stage is the hot gas stream at temperature $T_d^{in}(t)$, which represents the inlet gas conditions for the drying zone. Any additional air injected at this point can be used for fine-tuning oxygen content or adjusting gas temperature. A well-designed windbox ensures even gas distribution, preventing channeling or uneven drying across the pebble bed [17]–[19].

### C. Drying Zone (Pebble Bed)

The drying zone is the core of the process. It contains a bed of phosphate pebbles through which the hot combustion gases flow. Two major mass streams define this subsystem:

- Wet pebble feed: Pebbles enter the dryer at mass flow $F_s(t)$ with an inlet moisture fraction $X_{in}(t)$.
- Dried pebble product: Pebbles exit at the same mass flow $F_s(t)$ but with a reduced moisture content $X_{out}(t)$.

Within this zone, moisture is removed by two mechanisms:

- Convective heat transfers from the hot gases to the pebbles, raising their temperature.
- Evaporation of water inside the pores and surface of the pebbles, consuming latent heat.

The gas–solid contact is enhanced by the bed geometry and airflow distribution, ensuring efficient heat and mass transfer. The outlet solids define the primary product stream, while the outlet moisture level $X_{out}(t)$ is the critical quality variable of the process. The drying zone is also the principal location where disturbances (i.e., fluctuations in inlet moisture or solids feed) must be rejected through control action [20]–[23].

### D. Exhaust Stack and Induced-Draft Fan

The exhaust system removes the moisture-laden gases from the dryer. The gas stream leaving the bed contains combustion products and the evaporated water vapor. This flow is extracted through the stack by means of an induced draft (ID) fan.

The ID fan serves two critical roles:

- Maintaining a slight negative pressure inside the dryer to prevent backflow of gases into upstream equipment.
- Regulating the exhaust flowrate $\dot{m}_{stack}(t)$, which determines the residence time of gases in the system and affects drying efficiency.

The exhaust stream is characterized by the stack gas temperature and pressure, both of which are used for process monitoring and control. The stack may also be equipped with heat-recovery or pollution-control devices, but its primary function is safe discharge of the spent gases [24]–[26].

## III. System Modeling

The drying process of phosphate pebbles begins with the introduction of a pebble–water slurry into a dryer bed, where the material is directly contacted by hot combustion gases. Heat transfer between the gases and solids drives moisture evaporation, reducing the initial moisture content of about 15% by weight to approximately 3% at the dryer outlet. The primary control objective is to regulate the outlet moisture content $X_{out}(t)$ by manipulating the table-feeder speed, which sets the solid feed rate $F_s(t)$. While fuel flow is often used as a moisture control variable in similar dryers, in this process the feedrate is chosen as the main manipulated variable. Auxiliary loops are included to stabilize operation, such as combustion chamber temperature control via the fuel/air ratio, ID fan control to regulate vacuum pressure, and FD fan control for air supply management. The main disturbances to the system are variations in inlet slurry moisture and flowrate, which directly influence drying demand.

The system can be divided into four subsystems. The first is the combustion furnace, where fuel flow $\dot{m}_f(t)$, supplied by a pump, mixes with combustion air flow $\dot{m}_a(t)$, supplied by the FD-fan. Complete combustion generates hot flue gases, represented by mass $m_c(t)$ and temperature $T_c(t)$, which then flow to the windbox. The windbox serves as an air distribution plenum, equalizing the flow and stabilizing the gas temperature. Its state variables are gas mass $m_w(t)$ and temperature $T_w(t)$, while the outlet stream provides hot gases to the drying zone at temperature $T_d^{in}(t)$.

The drying zone contains the pebble bed, modeled with a fixed dry solid mass $M_s$, a time-varying water mass $M_w(t)$, the bed temperature $T_s(t)$, and the moisture fraction $X(t) = M_w(t)/(M_s(t) + M_w(t))$. Wet pebbles enter with feedrate $F_s(t)$ and inlet moisture $X_{in}(t)$, and dried pebbles exit with the same flowrate but reduced moisture $X_{out}(t)$. Simultaneously, evaporated moisture enters the gas stream. The gas in the drying zone is represented by mass flow $\dot{m}_g(t)$ and temperature $T_g(t)$, which is usually close to the windbox gas temperature unless the gas residence time in the bed becomes significant.

Finally, the exhaust subsystem collects the flue gases, characterized by mass $m_e(t)$ and $T_e(t)$ before they are released to the atmosphere through the stack. This is facilitated by the ID fan, which regulates the flow $\dot{m}_{stack}(t)$ and ensures the proper draft pressure inside the system. Across all subsystems, the gas phases are assumed to behave as ideal gases under constant pressure, while temperatures are considered perfectly mixed within each volume. Combustion is assumed complete, with instantaneous release of heat via the fuel's heating value, and the dominant mode of heat transfer from gas to solids is convection at the bed surface, modeled using an overall heat transfer coefficient $U$ and an effective surface area $A$.

### A. Combustion Furnace Modeling

In a gas turbine or furnace system, the combustion chamber is a control volume where fuel and air react to produce hot flue gases. In modeling this chamber as a lumped volume, one defines the fuel mass flow $\dot{m}_f(t)$ and the air mass flow $\dot{m}_a(t)$ entering the chamber, and the chamber's total gas mass $m_c(t)$ at temperature $T_c(t)$. By the conservation of mass, the gas mass inside the chamber changes only by the difference between inflows and outflow. In particular, complete combustion is assumed so that all fuel and air become flue gas; thus, the mass flow out of the chamber (toward the windbox or turbine) is the sum of the fuel and air flows [27]. The lumped mass balance can be written as:

$$\frac{\mathrm{d}m_c(t)}{\mathrm{d}t} = \dot{m}_f(t) + \dot{m}_a(t) - \dot{m}_{c\to w}(t) \tag{1}$$

where $\dot{m}_{c\to w}(t)$ is the mass flow of combustion gases leaving the chamber into the downstream component [27]. It can also see that in unsteady operation, $m_c(t)$ can rise or fall, corresponding to pressure or density changes in the chamber.

The first law of thermodynamics (energy conservation) applied to the chamber gives an enthalpy balance on the gas. Neglecting heat losses to the chamber walls (adiabatic assumption) and kinetic/potential energy changes, the rate of change of the chamber's total enthalpy $C_p^c\, m_c(t)\, T_c(t)$ equals the enthalpy inflow plus fuel heat release minus the enthalpy carried out. Mathematically, one can write:

$$C_p^c \cdot \frac{\mathrm{d}\big(m_c(t)\cdot T_c(t)\big)}{\mathrm{d}t} = HV\cdot\dot{m}_f(t) + C_p^a\cdot\dot{m}_a(t)\cdot(T_a(t) - \\ -T_c(t) - C_p^c\cdot\dot{m}_{c\to w}(t)\cdot\big(T_c(t) - T_w(t)\big) \tag{2}$$

where $HV$ is the fuel's heating value (energy released per unit mass of fuel), $C_p^c$ is the specific heat of the chamber gas mixture, $C_p^a$ is the air specific heat, $T_a(t)$ is the inlet air temperature, and $T_w(t)$ is the temperature of the gas leaving toward the windbox. In this expression, $HV\cdot\dot{m}_f(t)$ represents the thermal power input from fuel combustion (all fuel energy is assumed to heat the gas) [27]. The term $C_p^a\cdot\dot{m}_a(t)\cdot\big(T_a(t) - T_c(t)\big)$ accounts for the enthalpy carried in by the incoming air (if the air is hotter than the chamber, it adds enthalpy). The outflow term $C_p^c\cdot\dot{m}_{c\to w}(t)\cdot\big(T_c(t) - T_w(t)\big)$ is the enthalpy carried away by the exhaust gas as it leaves at temperature $T_w(t)$.

These equations (1)–(2) thus comprise a zero-dimensional dynamic model of the combustion chamber. In many gasses turbine models the chamber is assumed to operate at essentially constant pressure (isobaric combustion) [thermos-something], regulated by the surrounding flow or a draft/fan, so that pressure dynamics are not treated explicitly. Under this constant-pressure assumption (often used in Brayton-cycle analysis), the fuel heating value directly manifests as an increase in the gas enthalpy [thermos-something]. The outflow $\dot{m}_{c\to w}(t)$ from the chamber then feeds into the next component (for example, the windbox or the turbine inlet) in the overall system model.

## B. *Windbox Modeling*

In many thermal systems the windbox (or air distribution plenum) is a chamber that receives hot combustion gases (and any added air) and delivers them uniformly into a downstream bed or furnace. It acts as a mixing volume to equalize pressure and temperature, ensuring a stable, well-distributed flow. For example, in a fluidized-bed dryer hot air flows from the combustion chamber into the air distribution plenum, then through a perforated plate or tuyeres to evenly fluidize the particle bed. Likewise, in large boilers the windbox is a manifold that directs combustion air into the furnace in a way that creates turbulence and mixes it thoroughly with fuel. If flue gas recirculation is used, a separate "mixing plenum" upstream blends the recirculated flue gas with fresh air so that the windbox sees a homogeneous mixture [28]. In that case the windbox contains the combined stream – for instance, one analysis notes that "the windbox itself contains a mixture of air and wet flue gas comprising an intermediate oxygen fraction" [28] – reflecting that its contents are a blend of the inputs.

$$\frac{\mathrm{d}m_w(t)}{\mathrm{d}t} = \dot{m}_{c\to w}(t) - \dot{m}_{w\to d}(t) \tag{3}$$

where $\dot{m}_{w\to d}(t)$ is the gas mass flow into the drying zone. (If a blower supplies ambient air into the windbox as well, an inlet term $\dot{m}_w^{in}(t)$ would be added.). The energy balance for the windbox gas is:

$$C_p^w\cdot\frac{\mathrm{d}\big(m_w(t)\cdot T_w(t)\big)}{\mathrm{d}t} = C_p^c\cdot\dot{m}_{c\to w}(t)\cdot\big(T_c(t) - T_w(t)\big) \\ -C_p^w\cdot\dot{m}_{w\to d}(t)\cdot\big(T_w(t) - T_d^{in}(t)\big) \tag{4}$$

where $C_p^w$ is the gas specific heat in the windbox (similar to $C_p^c$), $T_c(t)$ is the temperature of gas entering from the chamber, and $T_d^{in}(t)$ is the gas temperature entering the dryer (in many cases $T_d^{in}(t) \approx T_w(t)$ if there is little time delay) [28]. The first term adds enthalpy from the incoming hot gas, and the second removes enthalpy with the flow into the dryer bed.

Expressions (3) and (4) allow $T_w(t)$ and $m_w(t)$ to respond dynamically if, for example, fuel or air flow changes cause changes in $T_c(t)$ or $\dot{m}_{c\to w}(t)$. The flow into the drying zone is $\dot{m}_{w\to d}(t)$ (for simplicity it is assumed $\dot{m}_{w\to d}(t) = \dot{m}_{c\to w}(t)$ if no losses).

## C. *Drying Zone Modeling*

The dryer contains a fixed bed of phosphate pebbles (dry mass $M_s(t)$) that move slowly through the bed. Let $M_w(t)$ be the total mass of water in the bed at time $t$, and $X(t) = M_w(t)/(M_s(t) + M_w(t))$ the wet-basis moisture fraction. The solid feed (table feeder) delivers pebbles at mass flow $\mathrm{d}M_s^{in}(t)/\mathrm{d}t = F_s(t)$ (dry solids) with inlet moisture $X_{in}(t)$ and removes dried pebbles at $\mathrm{d}M_s^{out}(t)/\mathrm{d}t = F_s(t)$ (assuming no solids buildup). The water mass flow in from the slurry is $\mathrm{d}M_w^{in}(t)/\mathrm{d}t = F_s(t)\, X_{in}(t)$, and water leaves the bed either as part of the removed solids ($F_s(t)\, X_{out}(t)$) or by evaporation into the gas [9]. Mass balance on the bed water yields:

$$\frac{\mathrm{d}M_w(t)}{\mathrm{d}t} = F_s(t)\cdot X_{in}(t) - F_s(t)\cdot X_{out}(t) \tag{5}$$

where $X_{out}(t)$ is the moisture of the solids exiting the bed. In a well-mixed (CSTR-like) approximation, the bed moisture equals the exit moisture, $X_{out}(t) = X(t) = M_w(t)/(M_s(t) + M_w(t))$. (Thus, the model ties the bed water mass $M_w(t)$ to the outlet moisture – controlling $F_s(t)$ changes $M_w(t)$ dynamics and hence $X_{out}(t)$.) The evaporation rate is $E(t) = F_s(t)\, X_{in}(t) - F_s(t)\, X_{out}(t)$ [9].

For energy, a heat balance on the wet bed (dry solid plus liquid water) is described. The bed temperature $T_s(t)$ is assumed uniform (average). The total thermal capacity of bed is $C_p^s\cdot M_s(t) + C_p^w\cdot M_w(t)$ (solid heat capacity plus liquid water) [9]. Heat is gained from the hot gas flowing through the bed and lost by the outflow of solids and evaporation of water. A concise form of the energy balance is:

$$\left(C_p^s \cdot M_s(t) + C_p^w \cdot M_w(t)\right) \cdot \frac{dT_s(t)}{dt} = U \cdot A \cdot \left(T_g(t) - T_s(t)\right) - LH \cdot \left(F_s(t) \cdot X_{in}(t) - F_s(t) \cdot X_{out}(t)\right) - -F_s(t) \cdot \left(C_p^s \cdot \left(T_s(t) - T_s^{in}(t)\right)\right) + +C_p^s \cdot X_{out}(t) \cdot \left(T_s(t) - T_w^{in}(t)\right) \quad (6)$$

where:

- $U \cdot A \cdot (T_g(t) - T_s(t))$ is the convective heat transfer from gas to bed ($T_g(t)$ is the gas temperature in the bed, approximately $T_w(t)$ entering the bed).
- $LH$ is the latent heat of vaporization (per unit mass of water), so $LH \cdot (F_s(t) \cdot X_{in}(t) - F_s(t) \cdot X_{out}(t)$ accounts for the energy required to evaporate the mass of water $E(t)$.
- The bracketed term represents enthalpy carried out by the solids stream: dry solids leaving carry $C_p^s$ each at $T_s(t)$ (assumed leaving at bed temperature), and the portion of water leaving in the solids ($X_{out}(t)$ fraction) carries $C_p^w$ each; $T_s^{in}(t)$, $T_w^{in}(t)$ are inlet temperatures of solids and water (often ambient), respectively. If incoming solids and water are at ambient (reference) temperature, the $T_s(t) - T_{in}(t)$ terms simplify.

Equation (6) can be expanded or simplified depending on assumptions; for example, if the bed is large then $M_s(t)$ dominates, and one may neglect $C_p^w \cdot M_w(t)$ term [9]. For completeness, one could also write a gas-phase balance through the bed (though often gas flow is fast and quasi-steady). A gas-energy balance over a control volume covering the bed would be, as expressed as follows:

$$\left(C_p^s \cdot M_s(t) + C_p^w \cdot M_w(t)\right) \cdot \frac{dT_s(t)}{dt} = U \cdot A \cdot \left(T_g(t) - T_s(t)\right) - LH \cdot \left(F_s(t) \cdot X_{in}(t) - F_s(t) \cdot X_{out}(t)\right) - -F_s(t) \cdot \left(C_p^s \cdot \left(T_s(t) - T_s^{in}(t)\right)\right) + +C_p^s \cdot X_{out}(t) \cdot \left(T_s(t) - T_w^{in}(t)\right) \quad (7)$$

From here, $\dot{m}_g^{out}(t)$ is the gas flow out of the bed (nearly equal to $\dot{m}_{w \to d}(t)$), and $T_e(t)$ is the exhaust gas temperature leaving the bed. In many analyses one assumes the gas enthalpy drop $-C_p^g \cdot \dot{m}_g^{out}(t) \cdot \left(T_g(t) - T_e(t)\right) - U \cdot A \cdot \left(T_g(t) - T_s(t)\right)$ accounts for heat to the bed and carries on [9].

## D. Exhaust System Modeling

Finally, the flue gases (now carrying most of the evaporated water vapor) exit through an exhaust duct or stack. Let $m_e(t)$ and $T_e(t)$ be the gas mass and temperature in this duct. The mass balance is:

$$\frac{dm_e(t)}{dt} = \dot{m}_g^{out}(t) - \dot{m}_{stack}(t) \quad (8)$$

where $\dot{m}_g^{out}(t)$ is the gas flow from the dryer and $\dot{m}_{stack}(t)$ is the flow leaving to the atmosphere (often controlled by a draft). For the energy balance, assuming ambient temperature $T_{amb}$ outside,

$$C_p^e \cdot \frac{d\left(T_e(t) \cdot m_e(t)\right)}{dt} = C_p^e \cdot \dot{m}_g^{out}(t) \cdot \left(T_d^{out}(t) - T_e(t)\right) - -C_p^e \cdot \dot{m}_{stack}(t) \cdot (T_e(t) - T_{amb}) - U_e \cdot A_e \cdot (T_e(t) - T_{amb}) \quad (9)$$

where $C_p^e$ is the flue gas specific heat, $T_d^{out}(t)$ is the gas temperature entering the exhaust (from the dryer), and $U_e$ $A_e$ represents any convective heat loss from the duct to the environment. (In many cases the exhaust volume is small, and the flow is almost immediately discharged, so $T_e(t) \approx T_d^{out}(t)$. Equations (8)–(9) capture dynamics such as startup transients of stack temperature or changes in gas flow due to fan action [29].

In addition to the governing mass and energy balances that characterize the exhaust subsystem, the dynamic behavior of the vacuum pressure induced by the ID fan is explicitly formulated [29]. The following expression provides a mathematical representation of the draft pressure ($P(t)$), linking it to the fan operation and the resulting gas flow conditions within the exhaust duct.

$$\frac{dP(t)}{dt} = \frac{R}{V_e} \cdot \frac{d\left(m_e(t) \cdot T_e(t)\right)}{dt} \quad (10)$$

After carrying out the required algebraic transformations, introducing the relevant constant definitions, and systematically expanding the product derivatives, the governing mass and energy balance equations were rigorously rearranged into synthetic dynamic equations. The resulting expressions provide a complete set of coupled nonlinear differential equations that capture the dynamic interactions of the drying system, thereby constituting the final nonlinear model of the process, as described as follows in (11).

## E. Steady-State Model

To obtain the linearized representation of the dryer system, it is first necessary to determine its steady-state operating points (OPs). These OPs define the equilibrium values of all process variables around which the linearization is performed. Linearization ensures that the resulting model accurately represents the system's behavior for small perturbations in the vicinity of these operating conditions [30].

By setting the time derivatives of the dynamic equations in (11) to zero, the nonlinear differential system is transformed into an algebraic system of equations that represents the steady-state mass and energy balances of all subsystems. Variables evaluated under steady conditions are denoted by the superscript "ss." The resulting system is given as follows in (12).

It must be emphasized that, for the mathematical formulation to remain physically consistent, the steady-state gas masses in the combustion chamber and the windbox must satisfy $m_c^{ss} \neq 0$ and $m_w^{ss} \neq 0$, respectively. These conditions guarantee the validity of the mass and energy conservation equations, avoiding singularities or indeterminate behavior in the corresponding differential system. From a physical standpoint, this assumption is justified because both control volumes continuously contain a finite gas inventory during steady and transient operation of the dryer.

$$
\left\{
\begin{array}{l}
\dfrac{\mathrm{d}m_c(t)}{\mathrm{d}t} = \dot{m}_\mathrm{f}(t) + \dot{m}_\mathrm{a}(t) - \dot{m}_{\mathrm{c}\to\mathrm{w}}(t) \\
\dfrac{\mathrm{d}T_\mathrm{c}(t)}{\mathrm{d}t} = k_{12}\cdot\dfrac{\dot{m}_\mathrm{f}(t)}{m_\mathrm{c}(t)} - \dfrac{\dot{m}_\mathrm{f}(t)}{m_\mathrm{c}(t)}\cdot T_\mathrm{c}(t) + k_{22}\cdot\dfrac{\dot{m}_\mathrm{a}(t)}{m_\mathrm{c}(t)}\cdot\left(T_\mathrm{a}(t) - T_\mathrm{c}(t)\right) - \dfrac{\dot{m}_\mathrm{a}(t)}{m_\mathrm{c}(t)}\cdot T_\mathrm{c}(t) - \\
\quad - \dfrac{\dot{m}_{\mathrm{e}\to\mathrm{w}}(t)}{m_\mathrm{c}(t)}\cdot\left(T_\mathrm{c}(t) - T_\mathrm{w}(t)\right) \\
\dfrac{\mathrm{d}m_\mathrm{w}(t)}{\mathrm{d}t} = \dot{m}_{\mathrm{c}\to\mathrm{w}}(t) - \dot{m}_{\mathrm{w}\to\mathrm{d}}(t) \\
\dfrac{\mathrm{d}T_\mathrm{w}(t)}{\mathrm{d}t} = k_{14}\cdot\dfrac{\dot{m}_{\mathrm{c}\to\mathrm{w}}(t)}{m_\mathrm{w}(t)}\cdot\left(T_\mathrm{c}(t) - T_\mathrm{w}(t)\right) - k_{24}\cdot\dfrac{\dot{m}_{\mathrm{c}\to\mathrm{w}}(t)}{m_\mathrm{w}(t)}\cdot T_\mathrm{w}(t) - \\
\quad -k_{34}\cdot\dfrac{\dot{m}_{\mathrm{w}\to\mathrm{d}}(t)}{\mathrm{m}_\mathrm{w}(t)}\cdot\left(T_\mathrm{w}(t) - T_\mathrm{d}^\mathrm{in}(t)\right) + k_{44}\cdot\dfrac{\dot{m}_{\mathrm{w}\to\mathrm{d}}(t)}{m_\mathrm{w}(t)}\cdot T_\mathrm{w}(t) \\
\dfrac{\mathrm{d}M_\mathrm{w}(t)}{\mathrm{d}t} = F_\mathrm{s}(t)\cdot\left(X_\mathrm{in}(t) - X_\mathrm{out}(t)\right) \\
\dfrac{\mathrm{d}m_\mathrm{g}(t)}{\mathrm{d}t} = F_\mathrm{s}(t)\cdot\left(\mathrm{X}_\mathrm{in}(t) - X_\mathrm{out}(t)\right) - \dot{m}_\mathrm{g}^\mathrm{out}(t) \\
\dfrac{\mathrm{d}T_\mathrm{g}(t)}{\mathrm{d}t} = \dfrac{\dot{m}_{\mathrm{w}\to\mathrm{d}}(t)}{m_\mathrm{g}(t)}\cdot\left(T_\mathrm{w}(t) - T_\mathrm{g}(t)\right) - \dfrac{F_\mathrm{s}(t)\cdot\left(X_\mathrm{in}(t) - X_\mathrm{out}(t)\right)}{m_\mathrm{g}(t)}\cdot T_\mathrm{g}(t) - \\
\quad - \dfrac{\dot{m}_\mathrm{g}^\mathrm{out}(t)}{m_\mathrm{g}(t)}\cdot\left(T_\mathrm{g}(t) - T_\mathrm{e}(t)\right) + \dfrac{\dot{m}_\mathrm{g}^\mathrm{out}(t)}{m_\mathrm{g}(t)}\cdot T_\mathrm{g}(t) - k_{17}\cdot\dfrac{\left(T_\mathrm{g}(t) - T_\mathrm{s}(t)\right)}{m_\mathrm{g}(t)} \\
\dfrac{\mathrm{d}m_\mathrm{e}(t)}{\mathrm{d}t} = \dot{m}_\mathrm{g}^\mathrm{out}(t) - \dot{m}_\mathrm{stack}(t) \\
\dfrac{\mathrm{d}T_\mathrm{e}(t)}{\mathrm{dt}} = \dfrac{\dot{m}_\mathrm{g}^\mathrm{out}(t)}{m_\mathrm{e}(t)}\cdot\left(T_\mathrm{d}^\mathrm{out}(t) - T_\mathrm{e}(t)\right) - \dfrac{\dot{m}_\mathrm{g}^\mathrm{out}(t)}{m_\mathrm{e}(t)}\cdot T_\mathrm{e}(t) - \dfrac{\dot{m}_\mathrm{stack}(t)}{m_\mathrm{e}(t)}\cdot\left(T_\mathrm{e}(t) - T_\mathrm{amb}\right) + \\
\quad + \dfrac{\dot{m}_\mathrm{stack}(t)}{m_\mathrm{e}(t)}\cdot T_\mathrm{e}(t) - \dfrac{\dot{m}_\mathrm{stack}(t)}{m_\mathrm{e}(t)}\cdot\left(T_\mathrm{e}(t) - T_\mathrm{amb}\right) - k_{18}\cdot\dfrac{\left(T_\mathrm{d}^\mathrm{out}(t) - T_\mathrm{amb}\right)}{m_\mathrm{e}(t)} \\
\dfrac{\mathrm{d}P(t)}{\mathrm{d}t} = k_{19}\cdot T_\mathrm{e}(t)\cdot\left(\dot{m}_\mathrm{g}^\mathrm{out}(t) - \dot{m}_\mathrm{stack}(t)\right) + k_{19}\cdot\dot{m}_\mathrm{g}^\mathrm{out}(t)\cdot\left(T_\mathrm{d}^\mathrm{out}(t) - T_\mathrm{e}(t)\right) - \\
\quad -k_{19}\cdot\dot{m}_\mathrm{g}^\mathrm{out}(t)\cdot T_\mathrm{e}(t) - k_{19}\cdot\dot{m}_\mathrm{stack}(t)\cdot\left(T_\mathrm{e}(t) - T_\mathrm{amb}\right) + \dot{m}_\mathrm{stack}(t)\cdot T_\mathrm{e}(t) -
\end{array}
\right. \tag{11}
$$

$$
\left\{
\begin{array}{l}
\dot{m}_\mathrm{f}{}^\mathrm{ss} + \dot{m}_\mathrm{a}{}^\mathrm{ss} - \dot{m}_{\mathrm{c}\to\mathrm{w}}{}^\mathrm{ss} = 0 \\
k_{12}\cdot\dot{m}_\mathrm{f}{}^\mathrm{ss} - \dot{m}_\mathrm{f}{}^\mathrm{ss}\cdot T_\mathrm{c}{}^\mathrm{ss} + k_{22}\cdot\dot{m}_\mathrm{a}{}^\mathrm{ss}\cdot(T_\mathrm{a}{}^\mathrm{ss} - T_\mathrm{c}{}^\mathrm{ss}) - \dot{m}_\mathrm{a}{}^\mathrm{ss}\cdot T_\mathrm{c}{}^\mathrm{ss} - \dot{m}_{\mathrm{e}\to\mathrm{w}}{}^\mathrm{ss}\cdot(T_\mathrm{c}{}^\mathrm{ss} - T_\mathrm{w}{}^\mathrm{ss}) = 0 \\
\dot{m}_{\mathrm{c}\to\mathrm{w}}{}^\mathrm{ss} - \dot{m}_{\mathrm{w}\to\mathrm{d}}{}^\mathrm{ss} = 0 \\
k_{14}\cdot\dot{m}_{\mathrm{c}\to\mathrm{w}}{}^\mathrm{ss}\cdot(T_\mathrm{c}{}^\mathrm{ss} - T_\mathrm{w}{}^\mathrm{ss}) - k_{24}\cdot\dot{m}_{\mathrm{c}\to\mathrm{w}}{}^\mathrm{ss}\cdot T_\mathrm{w}{}^\mathrm{ss} - k_{34}\cdot\dot{m}_{\mathrm{w}\to\mathrm{d}}{}^\mathrm{ss}\cdot\left(T_\mathrm{w}{}^\mathrm{ss} - T_\mathrm{d}^{\mathrm{in}^\mathrm{ss}}\right) + \\
\quad + k_{44}\cdot\dot{m}_{\mathrm{w}\to\mathrm{d}}{}^\mathrm{ss}\cdot T_\mathrm{w}{}^\mathrm{ss} = 0 \\
F_\mathrm{s}{}^\mathrm{ss}\cdot(X_\mathrm{in}{}^\mathrm{ss} - X_\mathrm{out}{}^\mathrm{ss}) = 0 \\
F_\mathrm{s}{}^\mathrm{ss}\cdot(X_\mathrm{in}{}^\mathrm{ss} - X_\mathrm{out}{}^\mathrm{ss}) - \dot{m}_\mathrm{g}^{\mathrm{out}^\mathrm{ss}} = 0 \\
\dot{m}_{\mathrm{w}\to\mathrm{d}}{}^\mathrm{ss}\cdot\left(T_\mathrm{w}{}^\mathrm{ss} - T_\mathrm{g}{}^\mathrm{ss}\right) - F_\mathrm{s}{}^\mathrm{ss}\cdot(X_\mathrm{in}{}^\mathrm{ss} - X_\mathrm{out}{}^\mathrm{ss})\cdot T_\mathrm{g}{}^\mathrm{ss} - \\
\quad -\dot{m}_\mathrm{g}^{\mathrm{out}^\mathrm{ss}}\cdot\left(T_\mathrm{g}{}^\mathrm{ss} - T_\mathrm{e}{}^\mathrm{ss}\right) + \dot{m}_\mathrm{g}^{\mathrm{out}^\mathrm{ss}}\cdot T_\mathrm{g}{}^\mathrm{ss} - k_{17}\cdot\left(T_\mathrm{g}{}^\mathrm{ss} - T_\mathrm{s}{}^\mathrm{ss}\right) = 0 \\
\dot{m}_\mathrm{g}^{\mathrm{out}^\mathrm{ss}} - \dot{m}_\mathrm{stack}{}^\mathrm{ss} = 0 \\
\dot{m}_\mathrm{g}^{\mathrm{out}^\mathrm{ss}}\cdot\left(T_\mathrm{d}^{\mathrm{out}^\mathrm{ss}} - T_\mathrm{e}{}^\mathrm{ss}\right) - \dot{m}_\mathrm{g}^{\mathrm{out}^\mathrm{ss}}\cdot T_\mathrm{e}{}^\mathrm{ss} - \dot{m}_\mathrm{stack}{}^\mathrm{ss}\cdot(T_\mathrm{e}{}^\mathrm{ss} - T_\mathrm{amb}) + \\
\quad + \dot{m}_\mathrm{stack}{}^\mathrm{ss}\cdot T_\mathrm{e}{}^\mathrm{ss} - \dot{m}_\mathrm{stack}{}^\mathrm{ss}\cdot(T_\mathrm{e}{}^\mathrm{ss} - T_\mathrm{amb}) - k_{18}\cdot\left(T_\mathrm{d}^{\mathrm{out}^\mathrm{ss}} - T_\mathrm{amb}\right) = 0 \\
k_{19}\cdot T_\mathrm{e}{}^\mathrm{ss}\cdot\left(\dot{m}_\mathrm{g}^{\mathrm{out}^\mathrm{ss}} - \dot{m}_\mathrm{stack}{}^\mathrm{ss}\right) + k_{19}\cdot\dot{m}_\mathrm{g}^{\mathrm{out}^\mathrm{ss}}\cdot\left(T_\mathrm{d}^{\mathrm{out}^\mathrm{ss}} - T_\mathrm{e}{}^\mathrm{ss}\right) - \\
\quad -k_{19}\cdot\dot{m}_\mathrm{g}^{\mathrm{out}^\mathrm{ss}}\cdot T_\mathrm{e}{}^\mathrm{ss} - k_{19}\cdot\dot{m}_\mathrm{stack}{}^\mathrm{ss}\cdot(T_\mathrm{e}{}^\mathrm{ss} - T_\mathrm{amb}) + \dot{m}_\mathrm{stack}{}^\mathrm{ss}\cdot T_\mathrm{e}{}^\mathrm{ss} - \\
\quad -\dot{m}_\mathrm{stack}{}^\mathrm{ss}\cdot(T_\mathrm{e}{}^\mathrm{ss} - T_\mathrm{amb}) - k_{29}\cdot\left(T_\mathrm{d}^{\mathrm{out}^\mathrm{ss}} - T_\mathrm{amb}\right) = 0
\end{array}
\right. \tag{12}
$$

In order to solve (12), the variables are separated into two sets: known variables and unknown variables. The known variable vector is defined as: $\mathbf{kv} = [\dot{m}_f^{ss}\ \dot{m}_a^{ss}\ F_s^{ss}\ X_{in}^{ss}\ T_a^{ss}\ X_{out}^{ss}]$ representing the measurable or imposed process quantities such as fuel flow, air flow, solid feed rate, inlet and outlet moisture contents, and inlet air temperature. The unknown variable vector, which must be determined from the steady- state system, is given by: $\mathbf{u}_v = [\dot{m}_{c\to w}^{ss}\ \dot{m}_{w\to d}^{ss}\ T_c^{ss}\ T_w^{ss}\ \dot{m}_g^{out^{ss}}\ \dot{m}_{stack}^{ss}\ T_g^{ss}\ T_e^{ss}\ T_d^{out^{ss}}]$.

Symbolically, the vectors are defined as $\mathbf{k}_v \in \{\mathbb{R}^6\}$ and $\mathbf{u}_v \in \{\mathbb{R}^9\}$. Solving (12) for $\mathbf{u}_v$ yields the OPs of the dryer, which are subsequently used as the linearization reference. The OPs are defined in (13). The corresponding parameter constants $k_{ij}$, where $i \in \{1, 2, 3\}$ and $j \in \{2, 4, 7, 8, 9\}$ used in the model are defined in (14).

### F. Linear Model in State Space

Once the OPs have been established, the nonlinear model in (12) can be linearized to obtain a local dynamic approximation of the system suitable for control design and stability analysis. The linearization is performed by expanding the nonlinear differential equations in a first-order Taylor series around the steady-state values of all state, input, and output variables. Higher-order terms are neglected under the assumption that the system operates within a sufficiently small neighborhood of the steady state, where nonlinear effects can be approximated as locally linear functions [30].

At the equilibrium condition, the steady-state relationships between the principal process variables are expressed in (13). These relationships capture the algebraic interdependence among the steady-state mass and energy flows within each subsystem (combustion chamber, windbox, drying zone, and exhaust). The model parameters $k_{ij}$ are defined in (14) and represent heat-transfer coefficients, specific-heat ratios, and geometric constants derived from thermophysical properties.

To facilitate the linearization, each process variable is decomposed into its steady-state value and a time-varying deviation component. Let $x(t)$ denote any process variable and $X^{ss}$ its steady-state counterpart.

$$\begin{cases} \dot{m}_{c\to w}^{ss} = \dot{m}_f^{ss} + \dot{m}_a^{ss} \\ \dot{m}_{w\to d}^{ss} = \dot{m}_{c\to w}^{ss} \\ T_c^{ss} = \dfrac{k_{12}\cdot\dot{m}_f^{ss} + k_{22}\cdot\dot{m}_a^{ss}\cdot T_a^{ss}}{\dot{m}_{c\to w}^{ss} + k_{22}\cdot\dot{m}_a^{ss}} + \dfrac{\dot{m}_{c\to w}^{ss2}\cdot\left(k_{34}\cdot T_d^{in^{ss}}\cdot\dot{m}_{w\to d}^{ss} + k_{12}\cdot k_{14}\cdot\dot{m}_f^{ss}\right) + k_{22}\cdot\dot{m}_a^{ss}\cdot\left(k_{14}\cdot\dot{m}_{c\to w}^{ss2}\cdot T_a^{ss} + k_{34}\cdot T_d^{in^{ss}}\cdot\dot{m}_{w\to d}^{ss}\cdot\dot{m}_{c\to w}^{ss}\right)}{\left(\dot{m}_{c\to w}^{ss} + k_{22}\cdot\dot{m}_a^{ss}\right)\cdot\left(k_{34}\cdot\dot{m}_{c\to w}^{ss}\cdot\dot{m}_{w\to d}^{ss} - k_{24}\cdot\left(\dot{m}_{c\to w}^{ss2} + k_{22}\cdot\dot{m}_a^{ss}\cdot\dot{m}_{c\to w}^{ss}\right) + k_{14}\cdot k_{22}\cdot\dot{m}_a^{ss}\cdot\dot{m}_{c\to w}^{ss} + k_{22}\cdot k_{34}\cdot\dot{m}_a^{ss}\cdot\dot{m}_{w\to d}^{ss}\right)} \\ T_w^{ss} = \dfrac{\dot{m}_{c\to w}^{ss}\cdot\left(k_{34}\cdot T_d^{in^{ss}}\cdot\dot{m}_{w\to d}^{ss} + k_{12}\cdot k_{14}\cdot\dot{m}_f^{ss} + k_{14}\cdot k_{22}\cdot\dot{m}_a^{ss}\cdot T_a^{ss}\right)}{\dot{m}_{c\to w}^{ss}\cdot\left(k_{34}\cdot\dot{m}_{w\to d}^{ss} + k_{14}\cdot k_{22}\cdot\dot{m}_a^{ss} - k_{22}\cdot k_{24}\cdot\dot{m}_a^{ss}\right) + k_{22}\cdot k_{34}\cdot\dot{m}_a^{ss}\cdot\dot{m}_{w\to d}^{ss}} + \dfrac{k_{22}\cdot k_{34}\cdot T_d^{in^{ss}}\cdot\dot{m}_a^{ss}\cdot\dot{m}_{w\to d}^{ss}}{\dot{m}_{c\to w}^{ss}\cdot\left(k_{34}\cdot\dot{m}_{w\to d}^{ss} + k_{14}\cdot k_{22}\cdot\dot{m}_a^{ss} - k_{22}\cdot k_{24}\cdot\dot{m}_a^{ss}\right) + k_{22}\cdot k_{34}\cdot\dot{m}_a^{ss}\cdot\dot{m}_{w\to d}^{ss}} \\ \dot{m}_g^{out^{ss}} = F_s^{ss}\cdot\left(X_{in}^{ss} - X_{out}^{ss}\right) \\ \dot{m}_{stack}^{ss} = \dot{m}_g^{out^{ss}} \\ T_g^{ss} = \dfrac{k_{17}\cdot T_s^{ss} + T_{amb}\cdot\dot{m}_g^{out^{ss}} + T_w^{ss}\cdot\dot{m}_{w\to d}^{ss}}{k_{17} + \dot{m}_g^{out^{ss}} + \dot{m}_{w\to d}^{ss}} + \dfrac{\dot{m}_g^{out^{ss}}\cdot\left(T_{amb} - T_d^{out^{ss}}\right)\cdot\left(k_{18} - \dot{m}_g^{out^{ss}}\right)}{\left(\dot{m}_g^{out^{ss}} - \dot{m}_{stack}^{ss}\right)\cdot\left(k_{17} + \dot{m}_g^{out^{ss}} + \dot{m}_{w\to d}^{ss}\right)} \\ T_e^{ss} = T_d^{out^{ss}} + \dfrac{\left(k_{18} + \dot{m}_{stack}^{ss}\right)\cdot\left(T_{amb} - T_d^{out^{ss}}\right)}{\dot{m}_g^{out^{ss}} + \dot{m}_{stack}^{ss}} \\ T_d^{out^{ss}} = T_g^{ss} \end{cases} \tag{13}$$

$$\begin{cases} k_{12} = \dfrac{HV}{C_p^c} \\ k_{12} = \dfrac{C_p^a}{C_p^c} \end{cases}, \begin{cases} k_{14} = \dfrac{C_p^c}{C_p^w} \\ k_{24} = \dfrac{C_p^a}{C_p^w} \\ k_{34} = 1 \end{cases}, k_{17} = \dfrac{U\cdot A}{C_p^w}, \quad k_{18} = \dfrac{U_e\cdot A_e}{C_p^e}, \begin{cases} k_{19} = \dfrac{R}{V_e} \\ k_{29} = \dfrac{R\cdot U_e\cdot A_e}{V_e\cdot C_p^e} \end{cases} \tag{14}$$

The deviation variable is then defined as: $x^{dv}(t) = x(t) - X^{ss}$. By definition, $x^{dv}(t)$ represents the instantaneous perturbation from steady state. Under the small-signal assumption widely used in process control and linear systems theory [30], [31], the deviation magnitude satisfies $|x^{dv}(t)| \ll |X^{ss}|$. Consequently, all nonlinear terms can be linearized with respect to $x^{dv}(t)$, yielding a state-space model in deviation-variable form that accurately describes the local dynamics near the selected OP.

The linearized model can be expressed compactly in vector–matrix notation as:

$$\begin{cases} \dfrac{\mathrm{d}\mathbf{x}^{\mathbf{dv}}(\mathrm{t})}{\mathrm{d}t} = \mathbf{A}\cdot\mathbf{x}^{\mathbf{dv}}(t) + \mathbf{B}\cdot\mathbf{u}^{\mathbf{dv}}(t) \\ \mathbf{y}^{\mathbf{dv}}(t) = \mathbf{C}\cdot\mathbf{x}^{\mathbf{dv}}(t) + \mathbf{D}\cdot\mathbf{u}^{\mathbf{dv}}(t) \end{cases} \tag{15}$$

where $\mathbf{x}^{\mathbf{dv}}(t)$ is the state vector, $\mathbf{u}^{\mathbf{dv}}(t)$ is the input vector, and $\mathbf{y}^{\mathbf{dv}}(t)$ is the output vector. For the dryer system, these vectors are defined as follows: $\mathbf{x}^{\mathbf{dv}}(t) = [m_c^{dv}(t), T_c^{dv}(t), m_w^{dv}(\mathrm{t}), T_w^{dv}(t), M_w^{dv}(t), m_g^{dv}(t), T_g^{dv}(t), m_e^{dv}(t), T_e^{dv}(t), P^{dv}(t)]^{\mathrm{T}}$, $\mathbf{u}^{\mathbf{dv}}(t) = [\dot{m}_f^{dv}(t), \dot{m}_a^{dv}(t), \dot{m}_{c\to w}^{dv}(t), \dot{m}_{e\to w}^{dv}(t), T_a^{ss}(t), \dot{m}_{w\to d}^{dv}(t), T_d^{in^{dv}}(t), F_s^{dv}(t), X_{in}^{dv}(t), X_{out}^{dv}(t), \dot{m}_g^{out^{dv}}(t), T_s^{dv}(t), \dot{m}_{stack}^{dv}(t), T_d^{out^{dv}}(t)]^{\mathrm{T}}$, and $\mathbf{y}^{\mathbf{dv}}(t) = \mathbf{x}^{\mathbf{dv}}(t)$, indicating that the system outputs correspond directly to measurable state variables. Symbolically, $\{\mathbf{x}^{\mathbf{dv}}(t), \mathbf{y}^{\mathbf{dv}}(t)\} \in \{\mathbb{R}^{10}\}$ and $\mathbf{u}^{\mathbf{dv}}(t) \in \{\mathbb{R}^{14}\}$.

On the other hand, **A**, **B**, **C**, and **D** correspond to the state matrix, input matrix, output matrix, and direct transmission matrix, respectively. The matrices are defined in (16). From here, **I** and **0** denote the identity and zero matrices, respectively. Symbolically, $\{\mathbf{A}, \mathbf{C}\} \in \mathscr{M}_{10\mathrm{x}10}\{\mathrm{K}\}$ and $\{\mathbf{B}, \mathbf{D}\} \in \mathscr{M}_{10\mathrm{x}14}\{\mathrm{K}\}$. The **A** matrix contains the linearized coupling coefficients ($\alpha_n$ where $n \in \{1, 2,\ldots, 33\}$) describing the interactions among temperature, mass flow, and pressure states, while **B** defines the sensitivities of the system states to input variations. The $\alpha_n$ coefficients are defined as follows in (17).

In this study, all state variables correspond to measurable quantities within the dryer subsystem; thus, the relationship $\mathbf{y}^{\mathbf{dv}}(t) = \mathbf{x}^{\mathbf{dv}}(t)$ holds [33]–[35].

This linearized state-space representation provides a mathematically tractable framework for analyzing system stability, designing feedback controllers, and evaluating dynamic responses under perturbations of fuel flow, air flow, or feed-rate conditions.

# IV. Dryer Efficiency Expression

The thermal efficiency of the phosphate-pebble dryer is defined as the ratio between the useful energy employed for water evaporation within the drying zone and the total thermal energy supplied by the combustion subsystem. The expression of this is described in (18). This dimensionless performance index quantifies the system's ability to convert input heat into latent energy of vaporization, reflecting the degree of thermal utilization in the process [9], [20], [32]–[34].

## *A. Useful Energy for Moisture Evaporation*

The useful thermal energy corresponds to the rate at which moisture is evaporated from the pebble bed. Considering the water evaporation rate $E(t) = F_s(t)\cdot[X_{in}(t) - X_{out}(t)]$ and the latent heat of vaporization $LH$, the instantaneous useful power is expressed as in (19).

Equation (19) directly couples the dryer efficiency to the solids feedrate and the moisture difference between inlet and outlet streams. This formulation aligns with first-law efficiency definitions widely applied in convective and rotary drying systems [32], [35]–[37].

## *B. Total Energy Input to the Dryer*

The total energy supplied to the system, $Q_{input}(t)$, originates from fuel combustion. Neglecting wall losses in the furnace and assuming complete combustion, the rate of heat release is defined in (20). From here, $HV$ is the lower heating value of the fuel and $\dot{m}_f(t)$ is the instantaneous fuel mass flow rate. A fraction of this energy is lost through the exhaust stack as sensible heat in the outgoing flue gases, quantified as in (21). From here, $C_p^e$ representing the specific heat of the exhaust gases, $T_e(t)$ the stack temperature, and $T_{amb}$ the ambient temperature [38], [39].

## *C. Thermal Efficiency Expression*

By substituting (19)–(21) into (18) and after some algebraic

$$\mathbf{A} = \begin{bmatrix} 0 & 0 & 0 & 0 & 0 & 0 & 0 & 0 & 0 & 0 \\ \alpha_4 & -\alpha_5 & 0 & \alpha_7 & 0 & 0 & 0 & 0 & 0 & 0 \\ 0 & 0 & 0 & 0 & 0 & 0 & 0 & 0 & 0 & 0 \\ 0 & \alpha_{11} & \alpha_{10} & \alpha_{12} & 0 & 0 & 0 & 0 & 0 & 0 \\ 0 & 0 & 0 & 0 & 0 & 0 & 0 & 0 & 0 & 0 \\ 0 & 0 & 0 & 0 & 0 & 0 & 0 & 0 & 0 & 0 \\ 0 & 0 & 0 & \alpha_{19} & 0 & \alpha_{18} & -\alpha_{20} & 0 & \alpha_{21} & 0 \\ 0 & 0 & 0 & 0 & 0 & 0 & 0 & 0 & 0 & 0 \\ 0 & 0 & 0 & 0 & 0 & 0 & 0 & \alpha_{27} & -\alpha_{29} & 0 \\ 0 & 0 & 0 & 0 & 0 & 0 & 0 & 0 & -\alpha_{32} & 0 \end{bmatrix},$$

$$\mathbf{B} = \begin{bmatrix} 1 & 1 & -1 & 0 & 0 & 0 & 0 & 0 & 0 & 0 & 0 & 0 & 0 & 0 \\ \alpha_1 & \alpha_2 & 0 & -\alpha_3 & \alpha_6 & 0 & 0 & 0 & 0 & 0 & 0 & 0 & 0 & 0 \\ 0 & 0 & 1 & -1 & 0 & 0 & -1 & 0 & 0 & 0 & 0 & 0 & 0 & 0 \\ 0 & 0 & \alpha_8 & 0 & 0 & \alpha_9 & \alpha_{13} & 0 & 0 & 0 & 0 & 0 & 0 & 0 \\ 0 & 0 & 0 & 0 & 0 & 0 & 0 & \alpha_{14} & \alpha_{15} & -\alpha_{15} & 0 & 0 & 0 & 0 \\ 0 & 0 & 0 & 0 & 0 & 0 & 0 & \alpha_{16} & \alpha_{15} & -\alpha_{15} & -1 & 0 & 0 & 0 \\ 0 & 0 & 0 & 0 & 0 & \alpha_{16} & 0 & -\alpha_{23} & -\alpha_{24} & \alpha_{24} & \alpha_{17} & \alpha_{22} & 0 & 0 \\ 0 & 0 & 0 & 0 & 0 & 0 & 0 & 0 & 0 & 0 & 1 & 0 & -1 & 0 \\ 0 & 0 & 0 & 0 & 0 & 0 & 0 & 0 & 0 & 0 & \alpha_{25} & 0 & \alpha_{26} & \alpha_{28} \\ 0 & 0 & 0 & 0 & 0 & 0 & 0 & 0 & 0 & 0 & \alpha_{30} & 0 & \alpha_{31} & \alpha_{33} \end{bmatrix}, \tag{16}$$

$$
\begin{aligned}
&\alpha_1 = \frac{k_{12} - T_c{}^{ss}}{m_c{}^{ss}},\ \alpha_2 = \frac{k_{22}\cdot(T_a{}^{ss} - T_c{}^{ss}) - T_c{}^{ss}}{m_c{}^{ss}},\ \alpha_3 = -\frac{T_e{}^{ss} - T_c{}^{ss}}{m_c{}^{ss}},\\
&\alpha_4 = \frac{-k_{12}\cdot\dot{m}_f{}^{ss} + \dot{m}_f{}^{ss}\cdot T_c{}^{ss} - k_{22}\cdot\dot{m}_a{}^{ss}\cdot(T_a{}^{ss} - T_c{}^{ss}) + \dot{m}_a{}^{ss}\cdot T_c{}^{ss} + \dot{m}_{e\to w}{}^{ss}\cdot(T_c{}^{ss} - T_w{}^{ss})}{m_c{}^{ss2}},\\
&\alpha_5 = \frac{\dot{m}_f{}^{ss} + k_{22}\cdot\dot{m}_a{}^{ss} + \dot{m}_a{}^{ss} + \dot{m}_{e\to w}{}^{ss}}{m_c{}^{ss2}},\ \alpha_6 = \frac{k_{22}\cdot\dot{m}_a{}^{ss}}{m_c{}^{ss}},\ \alpha_7 = \frac{\dot{m}_{e\to w}{}^{ss}}{m_c{}^{ss}},\\
&\alpha_8 = \frac{k_{14}\cdot(T_c{}^{ss} - T_w{}^{ss}) - k_{24}\cdot T_w{}^{ss}}{m_w{}^{ss}},\ \alpha_9 = \frac{-k_{34}\cdot\left(T_w{}^{ss} - T_d^{in^{ss}}\right) + k_{24}\cdot T_w{}^{ss}}{m_w{}^{ss2}},\\
&\alpha_{10} = \frac{-k_{14}\cdot\dot{m}_{c\to w}{}^{ss}\cdot(T_c{}^{ss} - T_w{}^{ss}) + k_{24}\cdot\dot{m}_{c\to w}{}^{ss}\cdot T_w{}^{ss} + k_{34}\cdot\dot{m}_{w\to d}{}^{ss}\cdot\left(T_w{}^{ss} - T_d^{in^{ss}}\right)}{m_w{}^{ss2}} -\\
&-\frac{k_{24}\cdot\dot{m}_{w\to d}{}^{ss}\cdot T_w{}^{ss}}{m_w{}^{ss2}},\ \alpha_{11} = \frac{k_{14}\cdot\dot{m}_{c\to w}{}^{ss}}{m_w{}^{ss}},\ \alpha_{12} = \frac{-k_{14}\cdot\dot{m}_{c\to w}{}^{ss} - k_{24}\cdot\dot{m}_{c\to w}{}^{ss}}{m_w{}^{ss}} +\\
&+\frac{k_{24}\cdot\dot{m}_{w\to d}{}^{ss} - k_{34}\cdot\dot{m}_{w\to d}{}^{ss}}{m_w{}^{ss}},\ \alpha_{13} = \frac{k_{34}\cdot\dot{m}_{w\to d}{}^{ss}}{m_w{}^{ss}},\ \alpha_{14} = X_{in}{}^{ss} - X_{out}{}^{ss},\ \alpha_{15} = F_s{}^{ss},\\
&\alpha_{16} = \frac{T_w{}^{ss} - T_g{}^{ss}}{m_g{}^{ss}},\ \alpha_{17} = \frac{T_e{}^{ss}}{m_g{}^{ss}},\ \alpha_{18} = \frac{\dot{m}_{w\to d}{}^{ss}\cdot\left(T_w{}^{ss} - T_g{}^{ss}\right)}{m_g{}^{ss2}} +\\
&+\frac{F_s{}^{ss}\cdot(X_{in}{}^{ss} - X_{out}{}^{ss})\cdot T_g{}^{ss} - \dot{m}_g^{out^{ss}}\cdot T_e{}^{ss} + k_{17}\cdot\left(T_g{}^{ss} - T_s{}^{ss}\right)}{m_g{}^{ss2}},\ \alpha_{19} = \frac{\dot{m}_{w\to d}{}^{ss}}{m_g{}^{ss}},\\
&\alpha_{20} = \frac{\dot{m}_{w\to d}{}^{ss} + F_s{}^{ss}\cdot(X_{in}{}^{ss} - X_{out}{}^{ss}) + k_{17}}{m_g{}^{ss}},\ \alpha_{21} = \frac{\dot{m}_g^{out^{ss}}}{m_g{}^{ss}},\ \alpha_{22} = \frac{k_{17}}{m_g{}^{ss}},\\
&\alpha_{23} = \frac{T_g{}^{ss}\cdot(X_{in}{}^{ss} - X_{out}{}^{ss})}{m_g{}^{ss}},\ \alpha_{24} = \frac{F_s{}^{ss}\cdot T_g{}^{ss}}{m_g{}^{ss}},\ \alpha_{25} = \frac{T_d^{in^{ss}} - 2\cdot T_e{}^{ss}}{m_e{}^{ss}},\\
&\alpha_{26} = \frac{-T_e{}^{ss} + 2\cdot T_{amb}}{m_e{}^{ss}},\ \alpha_{27} = \frac{-\dot{m}_g^{out^{ss}}\cdot T_d^{in^{ss}} + 2\cdot T_e{}^{ss}\cdot\dot{m}_g^{out^{ss}} - 2\cdot T_{amb}\cdot\dot{m}_{stack}{}^{ss}}{\mathrm{m}_e{}^{ss2}} +\\
&+\frac{T_e{}^{ss}\cdot\dot{m}_{stack}{}^{ss} + k_{18}\cdot\left(T_d^{out^{ss}} - T_{amb}\right)}{m_e{}^{ss2}},\ \alpha_{28} = \frac{\dot{m}_g^{out^{ss}} - k_{18}}{m_e{}^{ss}},\ \alpha_{29} = 2\cdot\frac{\dot{m}_g^{out^{ss}}}{m_e{}^{ss}},\\
&\alpha_{30} = k_{19}\cdot\left(T_d^{out^{ss}} - T_e{}^{ss}\right),\ \alpha_{31} = -2\cdot k_{19}\cdot T_e{}^{ss} + (1 + k_{19})\cdot T_{amb},\\
&\alpha_{32} = k_{19}\cdot\left(2\cdot\dot{m}_{stack}{}^{ss} + \dot{m}_g^{out^{ss}}\right),\ \alpha_{33} = k_{19}\cdot\dot{m}_g^{out^{ss}} - k_{29}
\end{aligned}
\tag{17}
$$

$$\eta_d(t) = \frac{Q_{useful}(t)}{Q_{input}(t)} \tag{18}$$

$$Q_{useful}(t) = LH\cdot F_s(t)\cdot\left(X_{in}(t) - X_{out}(t)\right) \tag{19}$$

$$Q_{input}(t) = HV\cdot\dot{m}_f(t) \tag{20}$$

$$Q_{loss}(t) = C_p^e\cdot\dot{m}_{stack}(t)\cdot(T_e(t) - T_{amb}) \tag{21}$$

operations, the overall instantaneous dryer efficiency is obtained and defined in (22).

This equation represents a dynamic energy-efficiency model that explicitly relates the drying performance to the measurable

$$\eta_d(t) = \frac{LH\cdot F_s(t)\cdot\left(X_{in}(t) - X_{out}(t)\right)}{HV\cdot\dot{m}_f(t)}\cdot\left(1 - \frac{C_p^e\cdot\dot{m}_{stack}(t)\cdot(T_e(t) - T_{amb})}{HV\cdot\dot{m}_f(t)}\right) \tag{22}$$

process variables of the integrated system: solids feed rate, fuel consumption, moisture removal, and exhaust conditions. The last term in brackets accounts for convective and sensible losses through the exhaust duct, as modeled in (9). Under steady conditions, $\eta_d(t)$ converges to a stationary value $\eta_d{}^{ss}$ describing the baseline dryer performance.

## D. Simplified Temperature-Based Efficiency

In practical operation, the thermal efficiency is often approximated through temperature ratios when direct measurement of moisture content is difficult. Assuming constant gas mass flow and negligible external losses, the simplified form reads into the following [32], [34]:

$$\eta_d(t) \approx \frac{T_d^{in}(t) - T_e(t)}{T_d^{in}(t) - T_{amb}(t)} \quad (23)$$

Equation (23) indicates that higher exhaust temperatures $T_e(t)$ reduce efficiency, as a larger portion of the input heat leaves the dryer without contributing to moisture evaporation. Conversely, when $T_e(t)$ approaches the wet-bulb or dew-point temperature of the exhaust mixture, efficiency approaches unity. To analyze the influence of ambient temperature on dryer efficiency, the ambient temperature ($T_{amb}$) is modeled as a time-dependent function. This dynamic formulation allows the study of transient effects in which ($T_{amb}(t)$) varies over time, enabling the evaluation of how fluctuations in environmental conditions impact the instantaneous $\eta_d(t)$. By expressing $T_{amb}(t)$ as a function of time, the model captures both steady-state and dynamic responses of the system under changing external thermal conditions.

## E. Transient Efficiency Dynamics

Considering the non-steady operation of industrial dryers, the instantaneous efficiency can be treated as a dynamic state variable. Differentiating (22) yields:

$$\begin{aligned}\frac{d\eta_d(t)}{dt} = &\frac{LH}{HV \cdot \dot{m}_f(t)} \cdot \frac{d}{dt}\Big(F_s(t) \cdot \big(X_{in}(t) - X_{out}(t)\big)\Big) - \\ &- \frac{LH \cdot F_s(t) \cdot \big(X_{in}(t) - X_{out}(t)\big)}{HV \cdot \dot{m}_f(t)^2} \cdot \frac{d\dot{m}_f(t)}{dt} - \\ &- \frac{C_p^e}{HV} \cdot \frac{d}{dt}\left(\frac{\dot{m}_{stack}(t) \cdot (T_e(t) - T_{amb})}{\dot{m}_f(t)}\right)\end{aligned} \quad (24)$$

Equation (24) characterizes the time-varying response of the dryer efficiency to disturbances in feed flow, moisture content, or combustion conditions. During startup or load changes, the transient evolution of $T_e(t)$, $F_s(t)$, and $X_{out}(t)$ produces dynamic efficiency trajectories that can be simulated using the nonlinear model presented in Section 3.

# V. Efficiency Sensitivity Analysis

The $\eta_d(t)$ can be approximated by (23), which represents the ratio between the actual temperature drop of the drying air and the maximum possible drop between inlet and ambient conditions. This formulation provides a direct measure of how effectively the system utilizes the available thermal energy.

To evaluate how small variations in operating temperatures affect the efficiency, an analytical sensitivity analysis is carried out. This approach focuses on local sensitivities—partial derivatives with respect to each variable—and elasticities, or normalized sensitivities, which express the relative influence of each variable on $\eta_d(t)$.

## A. Local Sensitivity Derivation

For a system where the inlet air temperature ($T_d^{in}(t)$), exhaust air temperature ($T_e(t)$), and ambient temperature ($T_{amb}(t)$) are independent variables, the partial derivatives of $\eta_d(t)$ are in [40], [41]:

$$\begin{cases} \dfrac{\partial \eta_d(t)}{\partial T_d^{in}(t)} = \dfrac{T_e(t) - T_{amb}(t)}{\left(T_d^{in}(t) - T_{amb}(t)\right)^2} \\ \dfrac{\partial \eta_d(t)}{\partial T_e(t)} = -\dfrac{1}{T_d^{in}(t) - T_{amb}(t)} \\ \dfrac{\partial \eta_d(t)}{\partial T_{amb}(t)} = \dfrac{T_d^{in}(t) - T_e(t)}{T_d^{in}(t) - T_{amb}(t)} \approx \eta_d(t) \end{cases} \quad (25)$$

The sign of each derivative provides clear physical interpretation:

- Inlet temperature ($T_d^{in}(t)$): The derivative is positive, indicating that increasing $T_d^{in}(t)$ (while keeping $T_e(t)$ and $T_{amb}(t)$ constant) enhances the efficiency. This occurs because a higher inlet temperature increases the available thermal potential for moisture removal, aligning with observations in convective drying systems where increased inlet temperature improves drying performance up to product quality constraints.
- Higher exhaust temperatures reduce $\eta_d(t)$. This is consistent with energy efficiency theory, where a larger portion of thermal energy leaves the system as waste heat. Studies on thermal dryers have demonstrated that reducing exhaust air temperature, either through extended residence time or heat recovery, yields substantial gains in overall efficiency.
- Ambient temperature ($T_{amb}(t)$): The derivative is positive, showing that an increase in ambient temperature results in a higher efficiency ratio. A higher $T_{amb}(t)$ decreases the denominator ($T_d^{in}(t) - T_{amb}(t)$), thereby increasing $\eta_d(t)$. Although this effect is generally less pronounced, it suggests that preheating the inlet air or operating under warmer ambient conditions marginally enhances the apparent efficiency.

## B. Elasticity (Relative Sensitivity) Analysis

To quantify the relative impact of each variable, elasticities are derived by scaling the partial derivatives with the corresponding variable and efficiency value [40], [41] described in (26).

The elasticities satisfy the invariant relation $e_{T_d^{in}(t)} + e_{T_e(t)} + e_{T_{amb}(t)} = 0$, confirming that $\eta_d(t)$ is a ratio of temperature differences and therefore invariant under uniform scaling of all temperatures [40], [41].

From an operational standpoint:

- $e_{T_e(t)}$ typically exhibits the largest magnitude, confirming that the exhaust temperature exerts the strongest influence on efficiency. Small increases in $T_e(t)$ directly translate into significant efficiency losses.
- $e_{T_d^{in}(t)}$ is positive but of moderate magnitude, showing

$$
\begin{cases}
e_{T_d^{in}(t)}(t) = \dfrac{\partial \eta_d(t)}{\partial T_d^{in}(t)} \cdot \dfrac{T_d^{in}(t)}{\eta_d(t)} \Rightarrow e_{T_d^{in}(t)}(t) = \dfrac{T_d^{in}(t) \cdot \left(T_e(t) - T_{amb}(t)\right)}{\left(T_d^{in}(t) - T_{amb}(t)\right) \cdot \left(T_d^{in}(t) - T_e(t)\right)} \\
e_{T_e(t)}(t) = \dfrac{\partial \eta_d(t)}{\partial T_e(t)} \cdot \dfrac{T_e(t)}{\eta_d(t)} \Rightarrow e_{T_e(t)}(t) = -\dfrac{T_e(t)}{T_d^{in}(t) - T_e(t)} \\
e_{T_{amb}(t)}(t) = \dfrac{\partial \eta_d(t)}{\partial T_{amb}(t)} \cdot \dfrac{T_{amb}(t)}{\eta_d(t)} \Rightarrow e_{T_{amb}(t)}(t) = \dfrac{T_{amb}(t)}{T_d^{in}(t) - T_{amb}(t)}
\end{cases} \tag{26}
$$

that raising the inlet temperature increases $\eta_d(t)$, albeit with diminishing returns as $T_e(t)$ approaches $T_{amb}(t)$.

- $e_{T_{amb}(t)}$ remains the smallest in most cases, meaning that ambient fluctuations have a limited effect on performance compared with process temperatures.

## VI. Control System Analysis

For the analysis and design of the control system architecture, three decentralized single-input single-output (SISO) feedback loops are proposed to ensure stable and efficient operation of the dryer process. The controlled variables correspond to the outlet moisture content of the solids leaving the dryer bed $X_{out}^{dv}(t)$, the combustion chamber temperature $T_c^{dv}(t)$, and the draft pressure ($P^{dv}(t)$) generated by the ID fan. Each control loop is designed to regulate its corresponding process variable independently, based on the principle of input–output pairing and minimal dynamic interaction among loops [16], [42].

This multiloop configuration captures the essential dynamics of the drying process while maintaining implementation simplicity and robustness under load and feed disturbances. The inherent multivariable nature of the system—with strong coupling between heat and mass transfer phenomena—allows for multiple alternative loop configurations. However, the proposed decentralized structure provides a practical trade-off between control performance, loop independence, and operational feasibility. The overall control architecture is schematically represented in Fig. 2. The diagram in Fig. 2 illustrates three coordinated feedback control loops designed to regulate the key process variables of the drying system: $X_{out}(t)$, $T_c(t)$, and $P(t)$. Each controlled variable is paired with a dedicated compensator (Comp.) that generates its respective drive signal.

The first compensator regulates $X_{out}(t)$ by manipulating the solid feed rate deviation $F_s^{dv}(t)$. The second compensator acts on the combustion air flow through the FD fan, whose drive signal is denoted $c_2(t)$. This signal governs the air mass flow according to the dynamic relationship:

$$\dot{m}_a(t) = k_a \cdot c_2(t) \tag{27}$$

where $k_a$ represents the air-flow actuator gain.

The third compensator regulates the ID fan to maintain the system's draft pressure. Its control signal $c_3(t)$ drives the stack gas flow through the nonlinear relation:

$$\dot{m}_{stack}(t) = G_s \cdot c_3(t) \tag{28}$$

where the fan gain $G_s$ is defined as:

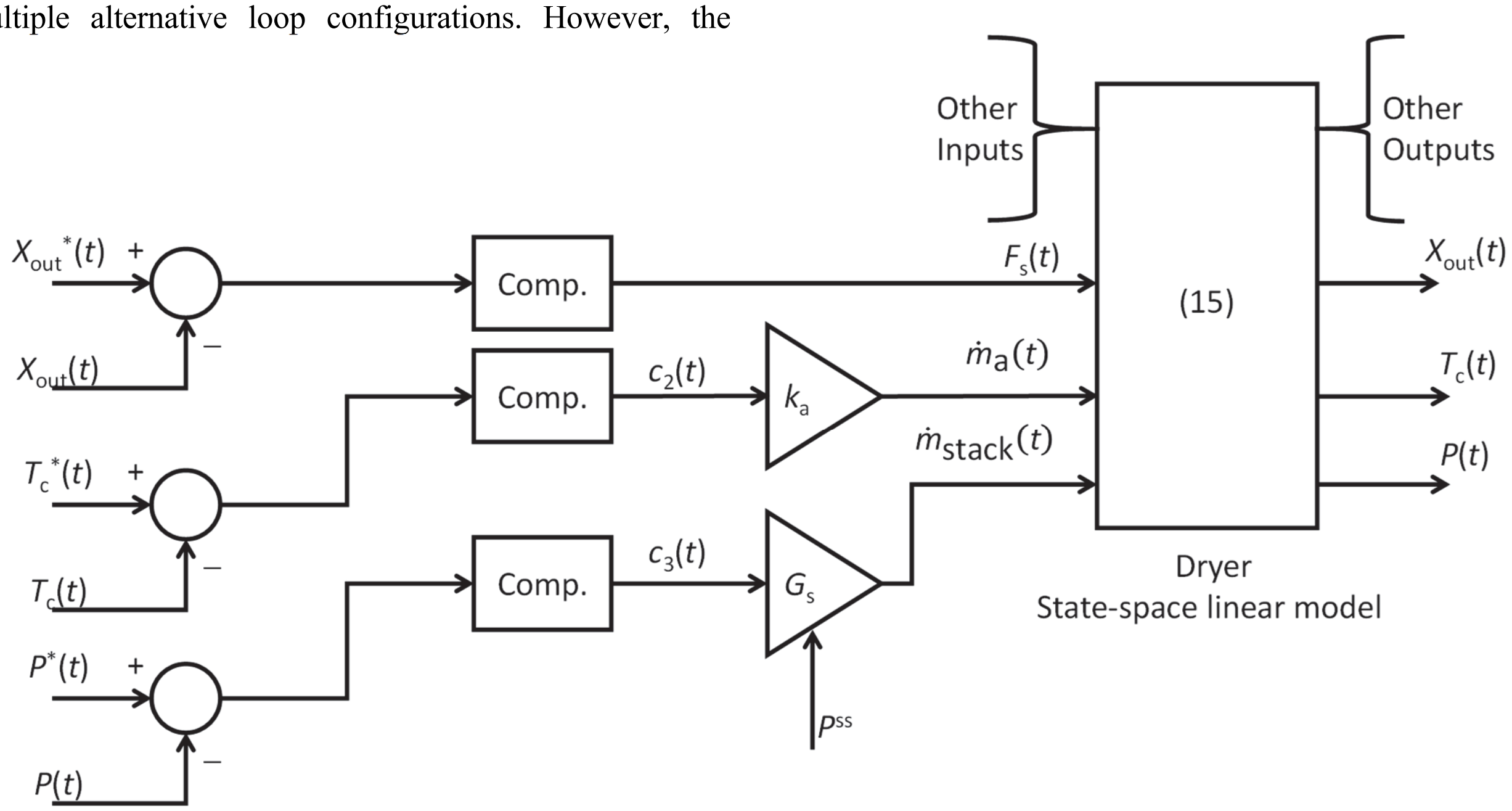

g. 2. Decentralized control architecture of the phosphate dryer. Three feedback loops regulate outlet moisture $X_{out}(t)$, chamber temperature $T_c(t)$, and draft pressure $P(t)$ using compensators acting on $F_s(t)$, $\dot{m}_a(t)$, and $\dot{m}_{stack}(t)$. The structure, based on (15), ensures robust multivariable control with minimal loop interaction.

$$G_s = k_f \cdot \sqrt{P^{ss}} \tag{29}$$

indicating that the exhaust flow depends on both the control signal and the square-root relationship between steady-state pressure and fan speed.

All three loops act upon the linearized process model represented by (15), which encapsulates the dryer's thermodynamic and hydrodynamic dynamics. The proposed configuration enables independent yet coordinated control of moisture removal, thermal input, and draft stabilization. This design minimizes loop interaction and enhances stability under coupled process disturbances, aligning with established principles of multivariable process control.

The subsequent section presents the detailed design of the control loops, focusing on the derivation and tuning of the compensator parameters for each feedback channel.

The control-loop design adopts direct synthesis (DS) within the internal model control (IMC) framework—a model-based strategy in which one first specifies a desired closed-loop response and then analytically inverts the process model (retaining non-invertible dynamics and adding a robustness filter) to obtain the controller. This procedure yields explicit compensator parameters tied to clear performance targets (i.e., time constants/bandwidth), provides built-in robustness and stability margins through the filter choice, and greatly reduces ad-hoc tuning effort compared with purely empirical methods[16]. This approach enables systematic tuning of compensators based on the process model, ensuring well-defined transient performance and stability margins while minimizing empirical adjustment.

### *A. Design of the Compensator Associated with $G_1(s) = X_{out}(s)/F_s(s)$ Loop*

This control loop regulates the moisture content of the dried solids, represented by the transfer function $G_1(s)$, which relates the manipulated variable $F_s(s)$ (the solids feed rate) to the controlled output $X_{out}(s)$ (the outlet moisture content). In the time-domain formulation of the state-space model (15), both variables appear as external inputs rather than explicit dynamic states. Consequently, it becomes necessary to derive an analytical relationship that dynamically couples $X_{out}(t)$ and $F_s(t)$, enabling the formulation of a tractable transfer function suitable for compensator design.

As recalled from Section 3.3, the instantaneous outlet moisture can be expressed as:

$$X_{out}(t) = M_w(t)/(M_s(t) + M_w(t)) \tag{30}$$

where $M_s(t)$ and $M_w(t)$ denote the dry-solid and water masses in the bed, respectively. The corresponding mass balance for the bed moisture is given by:

$$\frac{dM_w(t)}{dt} = F_s(t) \cdot X_{in}(t) - F_s(t) \cdot X_{out}(t) - E(t) \tag{31}$$

where $E(t)$ represents the instantaneous evaporation rate of water from the bed.

For the moisture-control loop, it is assumed that the gas-side dynamics (i.e., $T_g(t)$, $T_s(t)$) are tightly regulated by inner control loops, such that $E(t)$ varies only weakly with small perturbations in $F_s(t)$. Linearization is therefore performed around a steady-state operating point satisfying: $0 = F_s^{ss} \cdot [X_{in}^{ss} - X_{out}^{ss}] - E^{ss}$. Defining the deviation variables: $\delta x(t) = X_{out}(t) - X_{out}^{ss}$, $\delta f(t) = F_s(t) - F_s^{ss}$, and linearizing $X_{out}(t)$ with respect to $M_w(t)$, one obtains:

$$\left.\frac{dX_{out}(t)}{dM_w(t)}\right|_{ss} = \frac{M_s^{ss}}{(M_s^{ss} + M_w^{ss})^2} \Rightarrow$$
$$\Rightarrow \left.\frac{dX_{out}(t)}{dM_w(t)}\right|_{ss} = \frac{(1 - X_{out}^{ss})^2}{M_s^{ss}} \Rightarrow \left.\frac{dX_{out}(t)}{dM_w(t)}\right|_{ss} = k_x \tag{32}$$

where $k_x > 0$ denotes the steady-state sensitivity of moisture to bed water content.

Substituting (32) into the linearized moisture balance and neglecting perturbations in the evaporation term $\delta E(t)$ yields the first-order dynamic relation:

$$\frac{d\delta x(t)}{dt} = k_x \cdot (X_{in}^{ss} - X_{out}^{ss}) \cdot \delta f(t) - k_x \cdot F_s^{ss} \cdot \delta x(t) \tag{33}$$

Applying the Laplace transform (assuming zero initial conditions) leads to the TF in:

$$G_1(s) = X_{out}(s)/F_s(s) = K_1/(\tau \cdot s + 1) \tag{34}$$

From here, the process gain $K_1$ and time constant $\tau$ are obtained as indicated in:

$$\begin{cases} K_1 = k_x \cdot (X_{in}^{ss} - X_{out}^{ss}) \Rightarrow \\ \Rightarrow K_1 = \left(\dfrac{(1 - X_{out}^{ss})^2}{M_s^{ss}}\right) \cdot (X_{in}^{ss} - X_{out}^{ss}) \\ \tau = \dfrac{1}{k_x \cdot F_s^{ss}} \Rightarrow \tau = \dfrac{M_s^{ss}}{F_s^{ss} \cdot ((1 - X_{out}^{ss})^2)} \end{cases} \tag{35}$$

From these expressions, two physical insights arise.

First, $K_1 > 0$ since $X_{in}^{ss} > X_{out}^{ss}$; thus, an increase in solids feed rate raises the outlet moisture due to a reduced residence time.

Second, the process time constant $\tau$ decreases with increasing $F_s^{ss}$, implying faster moisture-response dynamics under higher throughputs—a behavior characteristic of convective drying systems.

These relationships establish the basis for controller synthesis and performance evaluation of the $X_{out}$-control loop [16], [30], [42].

Once the dynamic model of the drying subsystem has been characterized by the first-order transfer function (TF) in (34), the next step involves the design of the compensator (controller) that ensures satisfactory regulation of the outlet moisture content, $X_{out}(t)$, in response to variations in the feed rate $F_s(t)$ and external disturbances such as fluctuations in inlet moisture or gas temperature.

To this end, the DS method is adopted due to its transparency, analytical tractability, and direct link between desired closed-loop dynamics and the process model [16], [42], [43].

The DS method aims to shape the closed-loop response of the system according to a predefined desired dynamic behavior. For a SISO system described by a process TF) $G_p(s)$, the method begins by specifying a target closed-loop transfer function $G_{cl}^{des}(s)$, which reflects the desired performance in terms of settling time, rise time, and damping.

The general relationship between the compensator $G_c(s)$ and the process $G_p(s)$ in unity feedback configuration is expressed as:

$$G_{cl}(s) = G_c(s)\cdot G_p(s)/(1 + G_c(s)\cdot G_p(s)) \tag{36}$$

Imposing $G_{cl}(s) = G_{cl}^{des}(s)$, and solving for the compensator, yields the DS expression:

$$G_c(s) = G_{cl}^{des}(s)/(G_p(s)\cdot(1 - G_{cl}^{des}(s))) \tag{37}$$

Equation (37) provides a direct analytical relationship between the process dynamics, desired closed-loop dynamics, and the resulting compensator TF. The designer specifies $G_{cl}^{des}(s)$ to achieve desired time-domain or frequency-domain behavior while ensuring realizability and robustness.

For the drying process described by (34), the open-loop process transfer function is:

$$G_1(s) = K_1/(\tau\cdot s + 1) \tag{38}$$

where $K_1 > 0$ and $\tau$ are defined in (35).

The desired closed-loop response is typically chosen as a first-order system with a desired closed-loop time constant $\tau_c$, reflecting the target speed of response under a unity feedback configuration:

$$G_{cl}^{des}(s) = 1/(\tau_c\cdot s + 1) \tag{39}$$

Substituting (38) and (39) into the general DS formulation (37), the resulting compensator becomes:

$$G_{cl}(s) = (\tau\cdot s + 1)/(K_1\ \tau_c\cdot s) \tag{40}$$

Equation (40) defines a proportional-integral (PI) type controller, as it can be rewritten as:

$$G_{cl}(s) = (\tau/\ K_1\ \tau_c)\cdot(1 + 1/(\tau_c\cdot s)) \tag{41}$$

Grouping terms leads to the canonical PI form:

$$G_{cl}(s) = K_{cl}\cdot(1 + 1/(\tau_{I1}\cdot s)) \tag{42}$$

with:

$$K_{cl} = \tau/(\ K_1\ \tau_c),\ \tau_{I1} = \tau \tag{43}$$

*B. Design of the Compensator Associated with $G_2(s) = T_c(s)/\dot{m}_a(s)$ Loop*

The furnace–air loop regulates the combustion-chamber temperature $T_c(t)$ by manipulating the forced-draft air flow in (27). Around the steady state defined in Sections 3.5–3.6, the identified small-signal plant is:

$$G_2(s) = \frac{(\alpha_{12} - s)\cdot(\alpha_4 + \alpha_2\cdot s)}{-s^3 + (\alpha_{12} + \alpha_5)\cdot s^2 + (\alpha_5\cdot\alpha_{12} + \alpha_7\cdot\alpha_{11})\cdot s} \tag{44}$$

From (44), it can be seen the following implications:

- A pole at the origin comes from the factor s in the denominator (mass/energy accumulation in the chamber).
- A nonminimum-phase zero at $s = +\alpha_{12}$ (provided $\alpha_{12} > 0$, which occurs for the positive exchange/gain combination in (17)) captures the well-known "air addition initially cools $T_c(t)$" effect.
- An additional minimum-phase zero at $s = -\alpha_4/\alpha_2$ (typically left-half plane if $\alpha_2$, $\alpha_4 > 0$).

Neither the integrator nor the right-half plane (RHP) zero may be cancelled by the controller without violating internal stability or causality. Hence a DS/IMC design is adopted, which explicitly preserves non-invertible factors and regularizes the inversion with a low-pass filter [42], [44]–[48].

In order to develop a suitable controller, it can be factor $G_2$ into a non-invertible part $G_2^-$ and a minimum-phase part $G_2^+$ as follows:

$$\begin{cases} G_2^-(s) = \dfrac{\alpha_{12} - s}{s} \\ G_2^+(s) = \dfrac{\alpha_4 + \alpha_2\cdot s}{-s^2 + (\alpha_{12} + \alpha_5)\cdot s^2 + (\alpha_5\cdot\alpha_{12} + \alpha_7\cdot\alpha_{11})\cdot s} \\ G_2(s) = G_2^-(s)\cdot G_2^+(s) \end{cases} \tag{45}$$

Choose a strictly proper, stable IMC filter of order ≥ relative degree of $G_2^+$ (which is 1). A robust choice is the following:

$$F_2(s) = 1/(\lambda_2\cdot s + 1)^2 \text{ and } \lambda_2 > 0 \tag{46}$$

Equation (46) shapes bandwidth, caps high-frequency gain, and provides a single transparent tuning parameter $\lambda_2$ [44]–[46].

The IMC controller is defined as follows:

$$\begin{aligned} Q_2(s) &= G_2^+(s)^{-1}\cdot F_2(s) \Rightarrow \\ \Rightarrow Q_2(s) &= \frac{-s^2 + (\alpha_{12} + \alpha_5)\cdot s^2 + (\alpha_5\cdot\alpha_{12} + \alpha_7\cdot\alpha_{11})\cdot s}{\alpha_4 + \alpha_2\cdot s}\cdot \\ &\cdot\frac{1}{(\lambda_2\cdot s + 1)^2} \end{aligned} \tag{47}$$

Using the IMC–feedback equivalence for unity feedback with plant $G_2(s)$ [44]–[46] is described in:

$$\begin{aligned} G_{c2}(s) &= \frac{Q_2(s)}{1 - Q_2(s)\cdot G_2(s)} = \frac{G_2^+(s)^{-1}\cdot F_2(s)}{1 - G_2^-(s)\cdot F_2(s)} \Rightarrow \\ \Rightarrow G_{c2}(s) &= \frac{\dfrac{-s^2 + (\alpha_{12} + \alpha_5)\cdot s^2 + (\alpha_5\cdot\alpha_{12} + \alpha_7\cdot\alpha_{11})\cdot s}{\alpha_4 + \alpha_2\cdot s}\cdot}{1 - \dfrac{\alpha_{12} - s}{s\cdot(\lambda_2\cdot s + 1)^2}} \\ &\cdot\frac{1}{(\lambda_2\cdot s + 1)^2} \end{aligned} \tag{48}$$

From (48), this compensator does not cancel the integrator or the RHP and is strictly proper after rationalization. Also, it is tuned with the single knob $\lambda_2$.

The implemented drive compensator is $G_{c2}^{(drive)}(s) = G_{c2}(s)/k_a$ since the actuator signal satisfies (27).

Several key insights can be drawn from the compensator in (48), particularly regarding how the coefficients $\alpha_i$ in (17)

connect physical parameters to control performance:

- $\alpha_2$ increases with both $k_{22}$ and the air–to–chamber temperature difference ($T_a^{ss} - T_c^{ss}$); thus, stronger air enthalpy transfer enhances the process numerator slope and intensifies the furnace's thermal sensitivity to air-flow changes.
- $\alpha_4$ aggregates steady thermal flow contributions scaled by $(m_c^{ss})^{-2}$. A larger combustion-gas inventory $m_c^{ss}$ lowers $\alpha_4$, reducing dynamic gain and slowing numerator response.
- $\alpha_5$ represents the normalized total mass outflow; larger $\alpha_5$ increases damping in $G_2^+(s)$, providing greater thermal stability.
- $\alpha_{11} = k_{14}\cdot \dot{m}_{c\to w}^{ss}/m_w^{ss}$ and $\alpha_7 = \dot{m}_{e\to w}^{ss}/m_c^{ss}$ couple chamber and windbox energy transport. Their product $\alpha_7\cdot\alpha_{11}$ defines the constant term of the $G_2^+$ denominator, corresponding to the square of the natural frequency $\omega_n^2$.
- $\alpha_{12}$ defines both the RHP zero location through the term $(\alpha_{12} - s)$ and a denominator coefficient, establishing a fundamental limit on achievable control bandwidth. Increasing $\alpha_{12}$ shifts the zero further right, tightening the dynamic constraint of the furnace loop.

To select the filter time constant $\lambda_2$, the RHP zero at $+z = \alpha_{12}$ must be considered: attempting closed-loop poles much faster than $z$ produces severe undershoot and excessive actuation [16], [42], [47]–[50]. A conservative and robust guideline is therefore $\lambda_2 \geq 1/(2\cdot\alpha_{12})$.

The process's minimum-phase denominator is defined as follows:

$$D^+(s) = -s^2 + (\alpha_{12} - \alpha_5)\cdot s + (\alpha_7\cdot\alpha_{11} + \alpha_5\cdot\alpha_{12}) \tag{49}$$

From here, $\omega_n^2 = \alpha_7\cdot\alpha_{11} + \alpha_5\cdot\alpha_{12}$ defines the characteristic natural frequency. From this, the dominant dynamic time constant can be approximated as:

$$T_{dom} \approx \max\ \{1/\omega_n,\ 1/(\alpha_{12} - \alpha_5)_+\} \tag{50}$$

Hence, a robust and physically meaningful tuning for the IMC filter is:

$$\lambda_2 = \max\ \{1/\omega_n,\ 0.3\cdot T_{dom}\} \tag{51}$$

which can later be fine-tuned via frequency-domain loop-shaping of $L_2(s) = G_{c2}(s)\cdot G_2(s)$ to ensure a maximum sensitivity $M_s \leq 2$, phase margin $\geq 50°$, and gain margin $\geq 2$ [16], [42], [47]–[50].

### C. Design of the Compensator Associated with $G_3(s) = P(s)/\dot{m}_{stack}(s)$ Loop

This loop regulates the draft pressure $P(t)$ by manipulating the ID-fan exhaust flow $\dot{m}_{stack}(t)$. Linearization of the exhaust mass/energy balances (Sec. 3.4) yields the small-signal model:

$$G_3(s) = \frac{\alpha_{31}}{s} + \frac{\alpha_{32}\cdot(\alpha_{27} - \alpha_{26}\cdot s)}{s^2\cdot(\alpha_{29} + s)} \tag{52}$$

which is a double-integrating process with an additional first-order lag and a potential nonminimum-phase zero.

From (52), physics encoded in $\alpha$ − coefficients (from (17)) can be found, as follows:

- $\alpha_{31} = -(2\cdot k_{19})\cdot T_e^{ss} + (1 + k_{19})\cdot T_{amb}$: effective pressure–enthalpy gain ($k_{19} = R/V_e$), increasing with $T_{amb}$ and decreasing with $T_e^{ss}$; it weights the integrating term $1/s$ arising from compressible storage $dP(t)/dt = (R/V_e)\cdot d(m_e(t)\cdot T_e(t))/dt$.
- $\alpha_{32} = k_{19}\cdot(2\cdot \dot{m}_{stack}^{ss} + \dot{m}_g^{out^{ss}})$: flow–compressibility coupling; larger steady exhaust flow amplifies pressure sensitivity.
- $\alpha_{26} = (-T_e^{ss} + 2\cdot T_{amb})/m_e^{ss}$: thermal compressibility slope from exhaust energy balance.
- $\alpha_{27} = [-\dot{m}_g^{out^{ss}}\cdot T_d^{in^{ss}} + 2\cdot T_e^{ss}\cdot m\cdot\dot{m}_g^{out^{ss}} - 2\cdot T_{amb}\cdot\dot{m}_{stack}^{ss} + T_e^{ss}\cdot\dot{m}_{stack}^{ss} + k_{18}\cdot(T_d^{out^{ss}} - T_{amb})]/(m_e^{ss})^2$: steady enthalpy-flow mixture term.
- $\alpha_{29} = 2\cdot \dot{m}_g^{out^{ss}}/m_e^{ss}$: exhaust residence (inverse time constant) due to outflow; dominates the first-order lag $(\alpha_{29} + s)^{-1}$.

From the above, the process zero can be defined as follows: $z_3 = \alpha_{27}/\alpha_{26}$ valid for $\alpha_{26} \neq 0$. When $z_3 > 0$ the process contains an RHP zero $(1 - s/z_3)$, imposing a fundamental bandwidth limit and forbidding cancellation [16], [42], [44], [45], [47], [48].

With the aim of synthesizing the compensator, it can be written $G_3(s)$ as a strictly proper SISO model with separated non-invertible and invertible parts. Combine ($G_3(s)$) over the common denominator $s^2\cdot(\alpha_{29} + s)$:

$$G_3(s) = \frac{\alpha_{31}\cdot s\cdot(\alpha_{29} + s) + \alpha_{32}\cdot(\alpha_{27} - \alpha_{26}\cdot s)}{s^2\cdot(\alpha_{29} + s)} \tag{53}$$

Then, split:

$$G_3(s) = G_3^-(s)\cdot G_3^+(s) \tag{54}$$

with the expression in:

$$\begin{cases} G_3^-(s) = \dfrac{1}{s^2}\cdot\begin{cases}\left(1 - \dfrac{s}{z_3}\right), & z_3 > 0 \text{ (RHP zero, non-invertible)} \\ 1, & z_3 \leq 0\end{cases} \\ G_3^+(s) = \dfrac{N_3(s)}{(\alpha_{29}\ +\ s)}\cdot\begin{cases}\left(1 - \dfrac{s}{z_3}\right)^{-1}, & z_3 > 0 \\ 1, & z_3 \leq 0\end{cases}\end{cases} \tag{55}$$

So that $G_3^+(s)$ is minimum-phase and proper. Also, from (55) $N_3(s) = \alpha_{31}\cdot s\cdot(\alpha_{29} + s) + \alpha_{32}\cdot(\alpha_{27} + \alpha_{26}\cdot s)$.

In IMC, the internal controller is defined as follows:

$$\begin{cases} Q_3(s) = \left(G_3^+(s)\right)^{-1}\cdot F_3(s) \\ F_3(s) = \dfrac{1}{(\lambda_3\cdot s + 1)^n}\end{cases} \tag{56}$$

where $n \geq 2$ and the equivalent unity-feedback compensator is:

$$G_{c3}(s) = \frac{Q_3(s)}{1 - Q_3(s)\cdot G_3(s)} = \frac{G_3^+(s)^{-1}\cdot F_3(s)}{1 - G_3^-(s)\cdot F_3(s)} \tag{57}$$

To select the filter order, it is adopted—at least—a second-order filtering, i.e., $n = 2$ to ensure properness, noise attenuation, and robust roll-off [44], [45].

Substituting the explicit factors gives a realizable expression:

$$G_{c3}(s) = \frac{\dfrac{\alpha_{29} + s}{N_3(s)} \cdot \dfrac{1}{(\lambda_3 \cdot s + 1)^2}}{1 - \dfrac{1}{(\lambda_3 \cdot s + 1)^2} \cdot \dfrac{1}{s^2} \cdot \begin{cases} \left(1 - \dfrac{s}{z_3}\right), & z_3 > 0 \\ 1, & z_3 \leq 0 \end{cases}} \quad (58)$$

Equation (58) is the exact DS–IMC compensator that (i) does not cancel the RHP zero (if present), (ii) explicitly accounts for the double integration, and (iii) embeds a 2nd-order robustness filter.

## VII. Simulation Results

The nonlinear dynamic model of the dryer, governed by (11), was simulated under transient operating conditions using MATLAB/Simulink. The physical and thermodynamic parameters used in the simulation are provided in Table I.

To assess the closed-loop performance and dynamic response of the dryer system regulated by the designed compensators, several figures of merit (FoMs) were employed: the integral of squared error (*ISE*), the maximum percent overshoot (*ov*), and the final steady-state error (*ess*). The design criteria for satisfactory control performance specified a maximum overshoot of less than 20% and a steady-state error below 5%.

The simulation was conducted under non-cold-start conditions, with the system initialized from the steady-state values listed in Table II. These initial conditions were selected to facilitate a physically meaningful numerical integration of the system dynamics, avoiding artificial startup transients. The simulation sequence involved a series of carefully timed disturbances. At 200 seconds, a step decrease in fuel flow rate $\dot{m}_f(t)$ from 0.012 kg/s to 0.006 kg/s was applied. At 300 seconds, the burner outlet gas temperature $T_d^{in}(t)$ was increased from 720 °C to 792 °C. At 500 seconds, the inlet moisture content $X_{in}(t)$ rose from 15% to 23%, followed by a tightening of the outlet moisture specification $X_{out}(t)$ from 8% to 4% at 600 seconds. A substantial draft pressure disturbance was imposed at 800 seconds, doubling the vacuum level from −100 kPa to −200 kPa. Finally, the combustion chamber temperature $T_c(t)$ was increased from 800 °C to 1000 °C at 1,000 seconds. The simulation continued to 2,000 seconds to capture the system's full dynamic response.

TABLE I. DRYER SIMULATION PARAMETERS

| Variable | Value |
|---|---|
| Fuel flow ($\dot{m}_f(t)$) | 0.012 [kg/s] |
| Combustion/primary air ($\dot{m}_a(t)$) | 0.25 [kg/s] |
| Stack/exhaust outflow (ID fan $\dot{m}_{stack}(t)$) | 0.3 [kg/s] |
| Evaporated moisture flow ($\dot{m}_{e\to w}(t)$) | 0.2 [kg/s] |
| Combustion → windbox gas flow ($\dot{m}_{c\to w}(t)$) | 2 [kg/s] |
| Windbox → dryer gas flow ($\dot{m}_{w\to d}(t)$) | 1.8 [kg/s] |
| Exhaust gas flow ($\dot{m}_g^{out}(t)$) | 2 [kg/s] |
| Wet solids feed ($F_s(t)$) | 2.5 [kg/s] |
| Inlet moisture (wet basis $X_{in}(t)$) | 0.15 |
| Outlet moisture (wet basis $X_{out}(t)$) | 0.05 |
| Inlet gas to dryer (burner/windbox outlet $T_d^{in}(t)$) | 720 [°C] |
| Air temperature to combustion ($T_a(t)$) | 298 [°C] |
| Dryer outlet gas temperature ($T_d^{out}(t)$) | 370 [°C] |
| Gases entering windbox temperature ($T_w(t)$) | 900 [°C] |
| Ambient temperature ($T_{amb}$) | 293 [°C] |

TABLE II. INITIAL STATES USED IN SIMULATION

| Variable | Value |
|---|---|
| Initial combustion/windbox gas mass ($m_{c0}$) | 1.0 [kg] |
| Initial combustion/windbox gas temperature ($T_{c0}$) | 900 [°C] |
| Initial dryer gas mass ($m_{g0}$) | 2.0 [kg] |
| Initial exhaust gas mass ($m_{e0}$) | 2.0 [kg] |
| Initial dryer gas temperature ($T_{g0}$) | 420 [°C] |
| Initial exhaust gas temperature ($T_{e0}$) | 420 [°C] |
| Initial dry pebble bed mass ($M_{p0}$) | 750 [kg] |
| Initial pebble bed temperature ($T_{s0}$) | 370 [°C] |
| Initial moisture inventory within solids ($M_{w0}$) | 110 [kg] |
| Initial stack (draft) pressure ($P_0$) | −100 [kPa] |

Fig. 3 presents the closed-loop simulation results for the three main control loops: moisture ($G_1(s)$), combustion/windbox temperature ($G_2(s)$), and draft pressure ($G_3(s)$). In Fig. 3(a), the outlet moisture $X_{out}(t)$ (solid red) tracks its reference $X_{out}^*(t)$ (dashed blue), while the wet-solids feed $F_s(t)$ adjusts accordingly. $X_{out}(t)$ remains within control specifications during each disturbance, including those at 500 and 600 seconds. The performance meets the FoMs, with an overshoot below 20% and a steady-state error under 5%.

Fig. 3(b) displays the combustion temperature $T_c(t)$ (solid red), its reference $T_c^*(t)$ (dashed blue), the control signal $c_2(t)$, and the air flow rate $\dot{m}_a(t)$. Despite multiple disturbances, $T_c(t)$ remains within acceptable limits, quickly returning to its setpoint. In particular, the step at 1,000 seconds is managed with controlled rise and minimal overshoot.

Fig. 3(c) shows the draft pressure loop response: $P(t)$ (solid red) versus its reference $P^*(t)$ (dashed blue), the control signal $c_3(t)$, and the exhaust stack flow $\dot{m}_{stack}(t)$. Following the vacuum disturbance at 800 seconds, $P(t)$ undergoes only a small overshoot and returns to its reference within the design error band.

The calculated FoMs for these loops are summarized in Table III. The *ISE* values are minimal, and both *ov* and *ess* remain well within the design constraints, confirming excellent closed-loop performance.

Fig. 4 illustrates the drying system's behavior under undisturbed conditions. The control loops alone maintain key variables $X_{out}(t)$, $T_c(t)$, and $P(t)$ starting from their initial conditions. Initially, drying proceeds in the constant-rate regime: the bed is saturated with free moisture, and nearly all input energy is consumed as latent heat for evaporation. As drying transitions to the falling-rate phase, energy is increasingly spent on heating solids and gases, rather than moisture removal. As a result, drying efficiency $\eta_d(t)$ decreases from near 100% to a final value of approximately 35%, reflecting typical industrial dryer thermodynamic behavior. This trend aligns with heat and mass transfer theory and highlights the inherent inefficiency of thermal dryers in late-stage operation [23], [51].

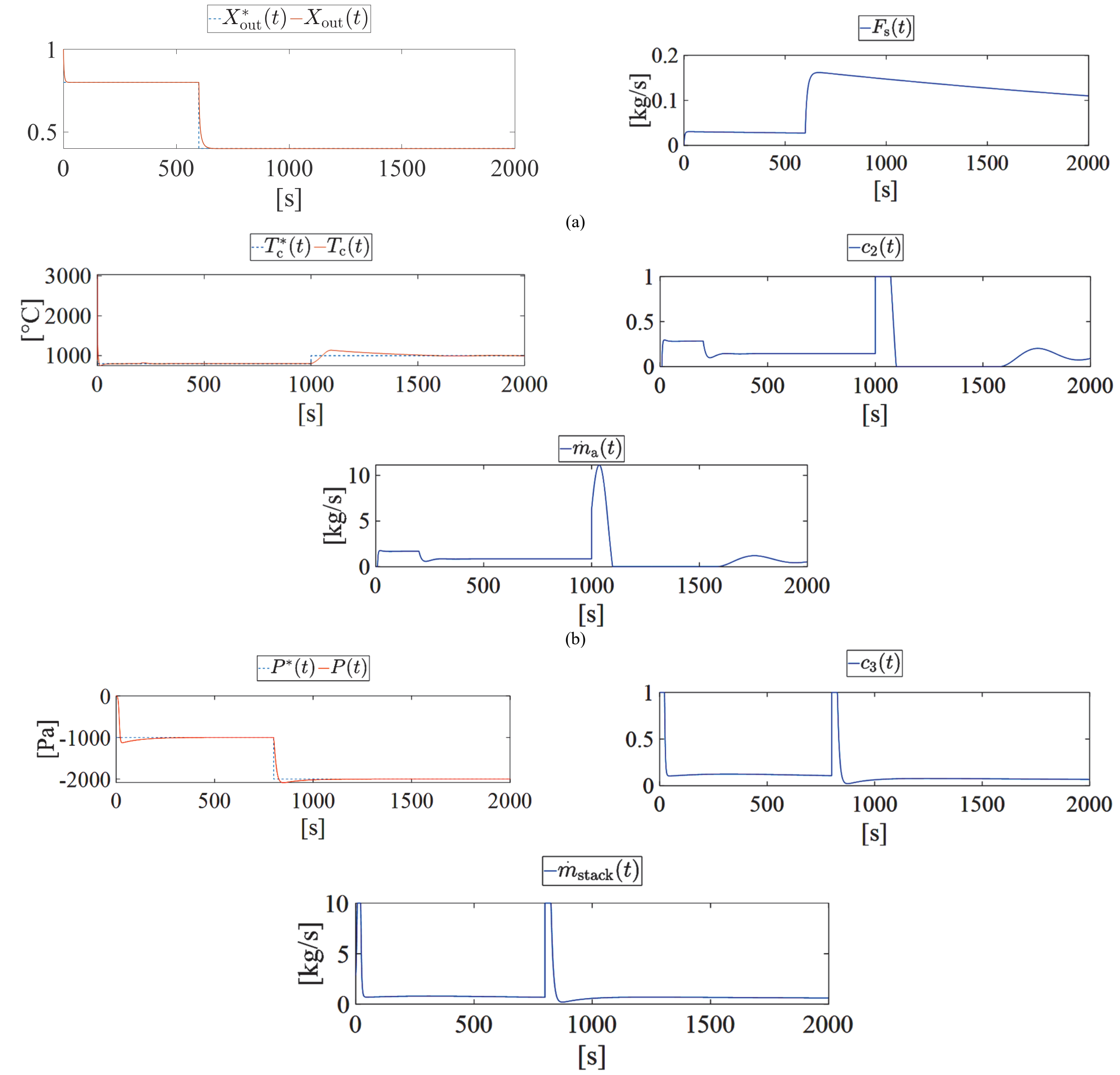

(c)

Fig. 3. Transient simulation results under non-cold-start conditions (initial states in Table II). Step disturbances: $\dot{m}_f(t)$ at 200 s, $T_d^{in}$ $(t)$ at 300 s, $X_{in}$ $(t)$ at 500 s, $X_{out}(t)$ at 600 s, $P(t)$ at 800 s, and $T_c(t)$ at 1,000 s. (a) $G_1(s)$ loop: $X_{out}(t)$, $F_s(t)$; (b) $G_2(s)$ loop: $T_c(t)$, $c_2(t)$, $\dot{m}_a(t)$; (c) $G_3(s)$ loop: $P(t)$, $c_3(t)$, $\dot{m}_{stack}(t)$.

TABLE III. OVERALL FOMS (WHOLE 0–2,000 S WINDOW)

| Loop | Overall *ISE* | Max *ov* [%] | Final *ess* [%] |
|---|---|---|---|
| Moisture — $X_{out}(t)$ | 0.021 | 12.4 | 3.1 |
| Temperature — $T_c(t)$ | $1.84\times10^5$ | 10.1 | 1.6 |
| Pressure — $P(t)$ | $3.10\times10^8$ | 14.7 | 2.3 |

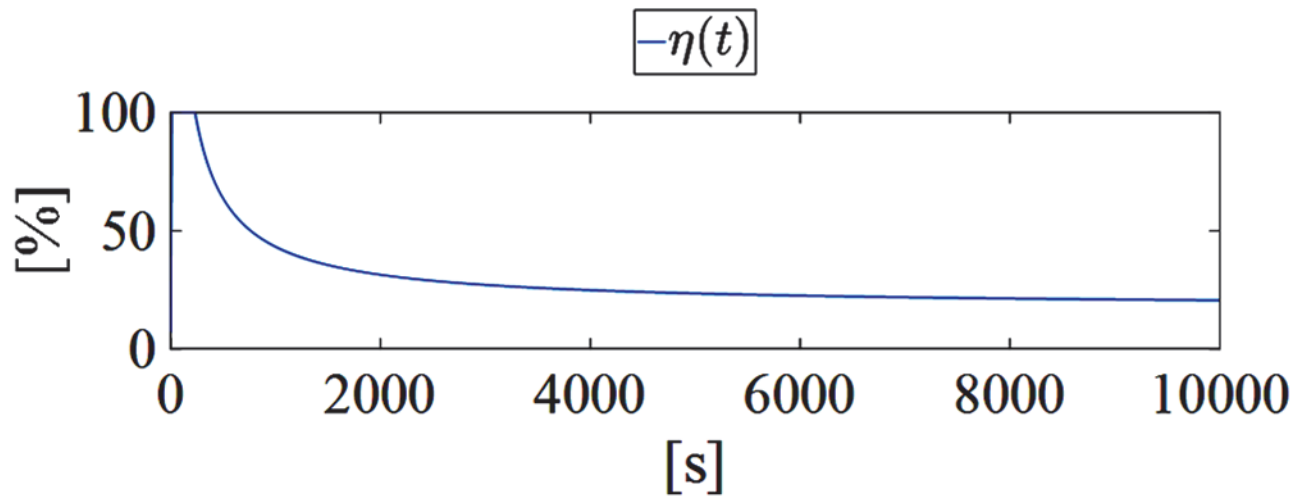

Fig. 4. Dryer performance under steady-state operation, where the system evolves from its initial conditions and the key variables $X_{out}(t)$, $T_c(t)$, and $P(t)$ are maintained exclusively by their respective control loops.

Fig. 5(a)—(c) present surfaces of ηd(t) computed from (23) under three scenarios: fixed ambient temperature $T_{amb}$(t) (Fig. 5(a)), fixed exhaust temperature $T_e(t)$ (Fig. 5(b)), and fixed inlet

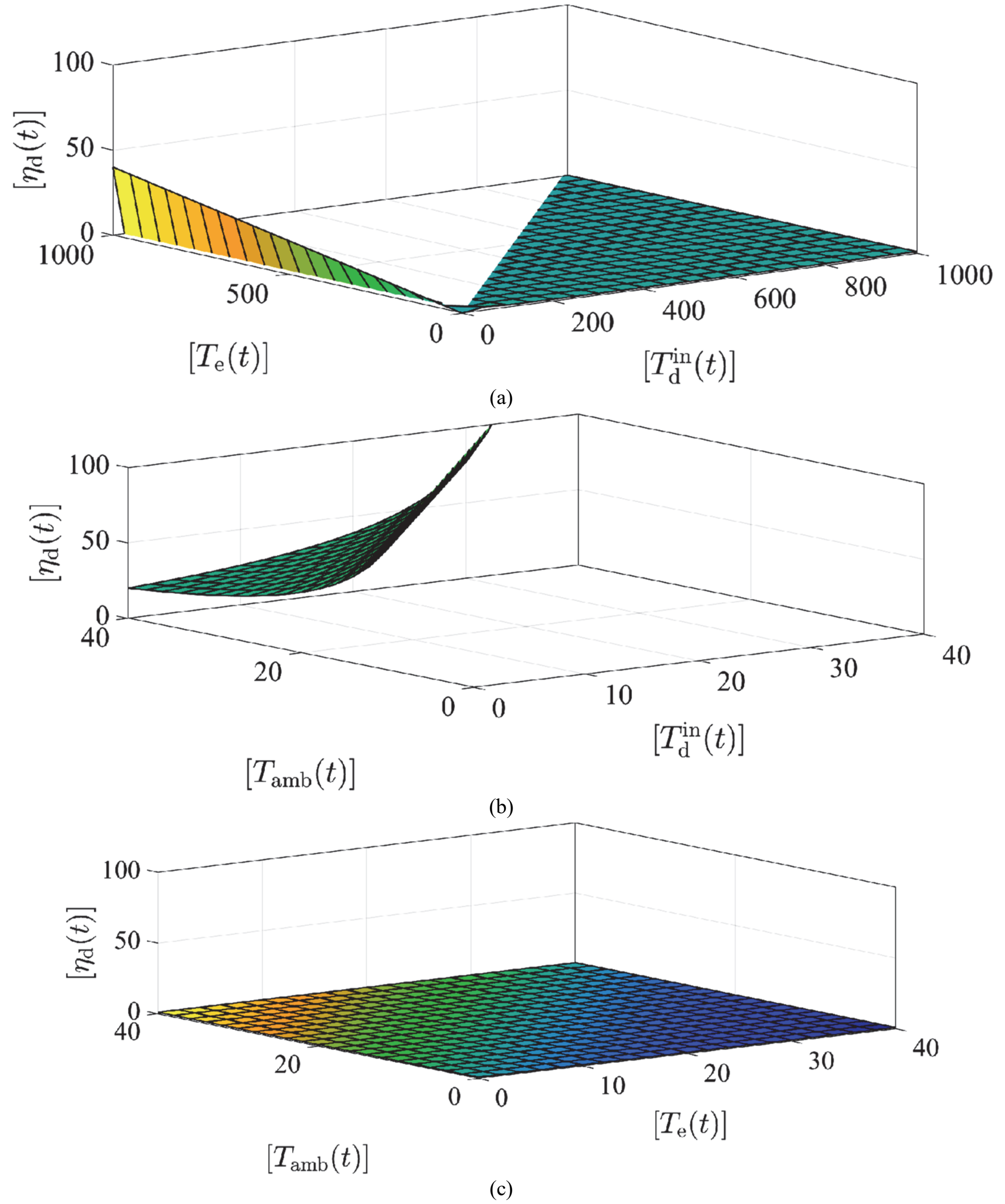

Fig. 5. Efficiency surfaces $\eta_d(t)$. (a) Scenario $\eta_d(t)|_{T_{amb}(t)=\text{constant}}$; (b) Scenario $\eta_d(t)|_{T_e(t)=\text{constant}}$; (c) Scenario $\eta_d(t)|_{T_{din}(t)=\text{constant}}$.

gas temperature $T_d^{in}(t)$ (Fig. 5(c)). In Fig. 5(a), efficiency increases with higher $T_d^{in}(t)$ and lower $T_e(t)$, since larger heat drops across the dryer correlate with greater moisture evaporation. When $T_e(t)$ approaches $T_d^{in}(t)$, efficiency vanishes due to lack of thermal driving force. This behavior reflects convective dryer limitations, where exhaust losses often account for 15–40% of input energy [9,24].

In Fig. 5(b), with fixed $T_e(t)$, efficiency increases with rising $T_d^{in}(t)$ and $T_{amb}(t)$. The model predicts higher $\eta_d(t)$ as $T_{amb}(t)$ increases, since this reduces the temperature lift (denominator), thereby increasing the apparent efficiency. However, elevated $T_{amb}(t)$ reduces drying potential. As shown in [52], [53], higher ambient conditions reduce irreversibility and increase nominal efficiency but may limit drying rates.

Fig. 5(c), holding $T_d^{in}(t)$ constant, highlights the inverse dependence of $\eta_d(t)$ on $T_e(t)$. As $T_e(t)$ increases, efficiency drops linearly; as $T_{amb}(t)$ increases, $\eta_d(t)$ increases. This illustrates the potential benefit of exhaust heat recovery and ambient air preheating to improve dryer performance [23], [54].

Fig. 6(a)—(c) plot sensitivity surfaces $\partial\eta_d(t)/\partial T_d^{in}(t)$. In Fig. 6(a), with $T_{amb}(t)$ constant, sensitivity increases as $T_e(t)$ decreases and $T_d^{in}(t)$increases. Fig. 6(b), with constant $T_e(t)$, shows peak sensitivity when $T_d^{in}(t)$ approaches $T_{amb}(t)$, indicating minimal enthalpic gain and poor dryer performance. Fig. 6(c), with constant $T_d^{in}(t)$, shows low overall sensitivity to $T_{amb}(t)$ variations, suggesting efficiencyrobustness to ambient fluctuations under high $T_d^{in}(t)$ operation.

Fig. 7 presents the sensitivity $\partial\eta_d(t)/\partial T_e(t)$. The surface is planar along the $T_e(t)$ axis, confirming the linear negative dependence of $\eta_d(t)$ on $T_e(t)$. Higher inlet temperatures $T_d^{in}(t)$ reduce this sensitivity, flattening the slope. This analytic structure is consistent with the form of equation (23) and with the theoretical basis of energy balance and exergy analysis in drying systems [23], [54].

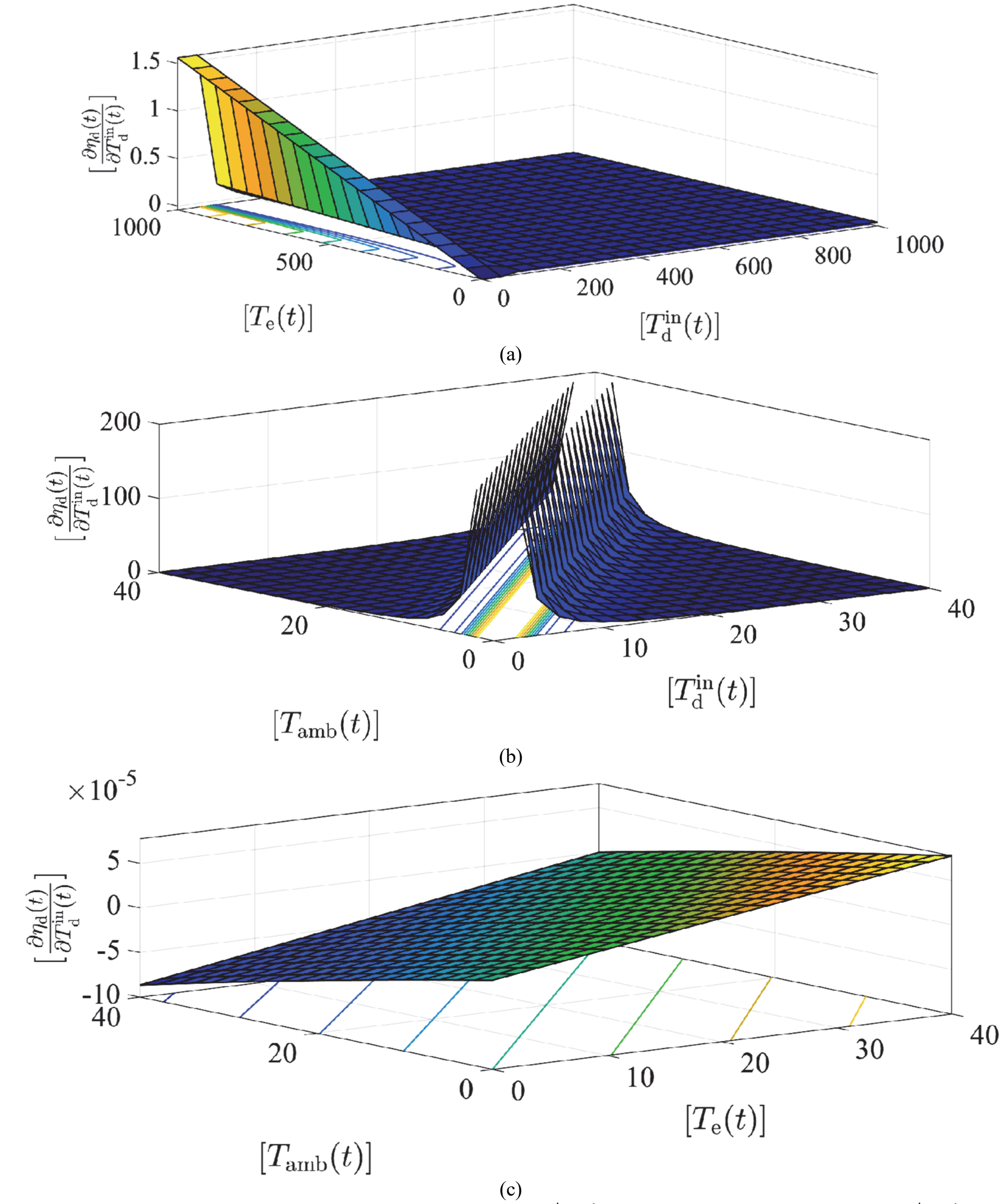

Fig. 6. Sensitivity surfaces of the drying efficiency $\eta_d(t)$. (a) Scenario $\partial\eta_d(t)/\partial T_d^{in}(t)|_{T_{amb}(t)=\text{constant}}$; (b) Scenario $\partial\eta_d(t)/\partial T_d^{in}(t)|_{T_e(t)=\text{constant}}$; (c) Scenario $\partial\eta_d(t)/\partial T_d^{in}(t)|_{T_d^{in}(t)=\text{constant}}$.

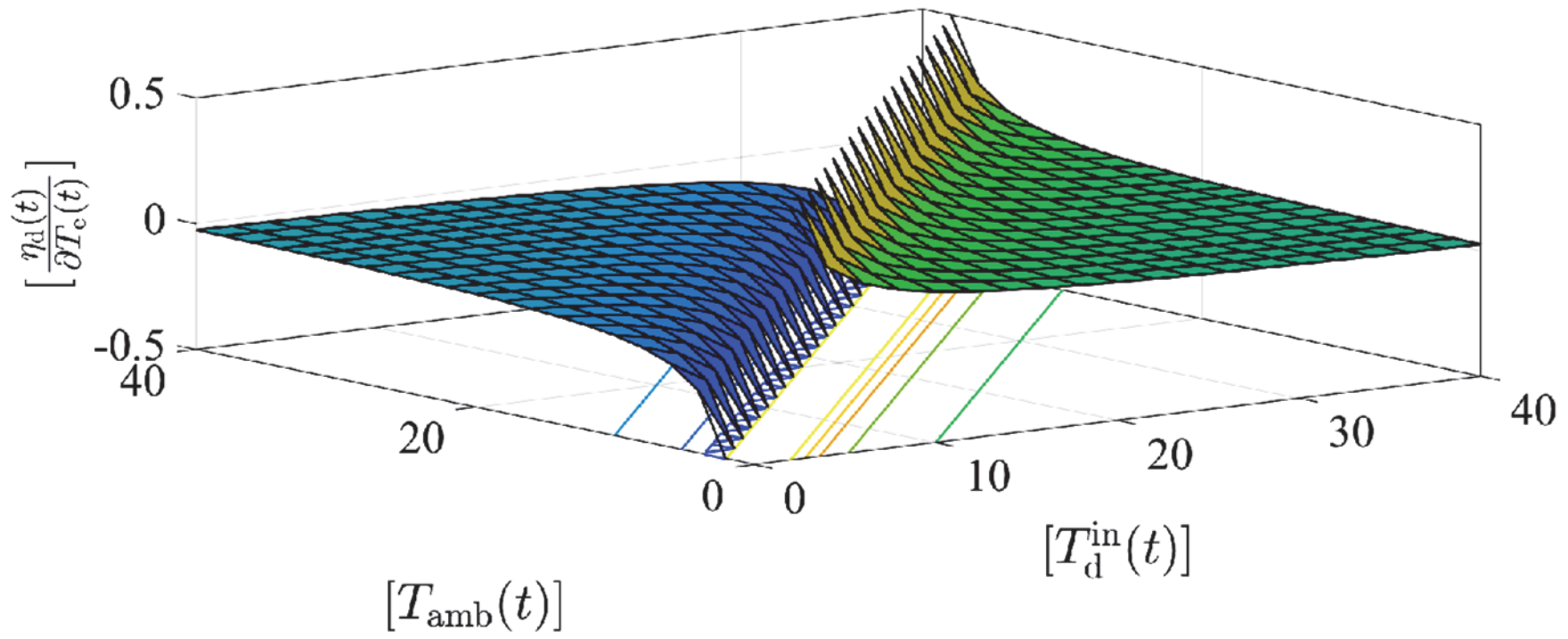

Fig. 7: Sensitivity surface of the drying efficiency $\eta_d(t)$ related with $\partial\eta_d(t)/\partial T_e(t)$.

Finally, the sensitivity of $\eta_d(t)$ to $T_{amb}(t)$ is analytically embedded in the model, as shown in Fig. 5(a)—(c). Since $T_{amb}(t)$ appears in the denominator of (23), its effect is nonlinear but bounded. Higher $T_{amb}(t)$ increases efficiency by shrinking the thermal lift, although at the cost of potential drying capacity. These effects reinforce the importance of controlling all three temperature boundaries—$T_d^{in}(t)$, $T_e(t)$, and $T_{amb}(t)$—to optimize energy utilization in convective dryers.

## VIII. CONCLUSION

This work presented a rigorous nonlinear dynamic modeling, simulation, and control analysis of a convective thermal dryer used for phosphate pebble drying. The dryer system, characterized by complex heat and mass transfer phenomena, was described via a high-fidelity nonlinear model derived from first principles and accounting for thermodynamic interactions among gas-phase flow, solid moisture evolution, and pressure dynamics. The model equations captured the transient behavior of inlet and exhaust gas temperatures, pressure drops, moisture content, and energy balances, integrating them into a unified framework implementable in MATLAB/Simulink.

Dynamic simulation scenarios under realistic operating disturbances—such as abrupt changes in fuel flow, inlet and outlet moisture content, draft pressure, and combustion chamber temperature—demonstrated the robustness and responsiveness of the system when operated under the proposed compensator structure. The three control loops—moisture content via wet feed rate, combustion temperature via air/fuel modulation, and draft pressure via exhaust flow—exhibited stable tracking with overshoot below 20%, steady-state error below 5%, and low *ISE* values, fulfilling the defined FoMs. These results confirmed the efficacy of the unit-gain feedback compensators in maintaining product quality and operational stability despite multiple input perturbations.

Furthermore, the analysis of drying efficiency $\eta_d(t)$, based on an interpretable algebraic formulation dependent on inlet, exhaust, and ambient temperatures, provided critical insights into thermodynamic performance. Simulations under undisturbed steady-state operation revealed that $\eta_d(t)$ begins near 100% during the constant-rate drying phase and progressively decreases to ~35% as the system transitions into the falling-rate regime, where energy is increasingly consumed as sensible heat and lost via exhaust gases and radiation. This efficiency profile aligns with thermophysical constraints and industrial drying behavior, highlighting the intrinsic limitations of single-stage convective drying systems.

The study also introduced and examined the efficiency surfaces and their sensitivity gradients with respect to temperature perturbations. Three-dimensional surface plots of $\eta_d(t)$ under varied thermodynamic configurations revealed the decisive role of thermal gradients in defining energy utilization. In particular, higher inlet temperatures and lower exhaust temperatures maximized efficiency, while warmer ambient conditions inflated apparent efficiency by reducing the baseline enthalpic lift. Sensitivity analyses further demonstrated that $\eta_d(t)$ is most sensitive to variations in inlet temperature $T_d^{in}(t)$ when exhaust or ambient temperatures are low—an insight critical for burner control and energy optimization strategies. Additionally, the linear and monotonic sensitivity of $\eta_d(t)$ with respect to exhaust temperature $T_e(t)$, and its inverse dependence on the inlet–ambient temperature difference, confirmed the analytical consistency and physical soundness of the proposed efficiency model.

In conclusion, this work not only validates a nonlinear dynamic model for industrial convective dryers but also provides a robust simulation framework and efficiency evaluation methodology. The findings underscore the importance of precise thermal management, real-time disturbance rejection, and targeted control design to improve energy efficiency and process stability. Future research may focus on incorporating advanced control strategies (i.e., model predictive control), implementing multi-stage or hybrid drying schemes, and integrating exhaust heat recovery systems to further enhance thermal performance and sustainability.